\documentclass[amsmath,aps,showpacs,a4paper,10pt]{revtex4}

 \usepackage{epsf}
 \usepackage{graphicx}    
 \usepackage{enumerate}   

\begin{document}

 \newcommand{\be}[1]{\begin{equation}\label{#1}}
 \newcommand{\ee}{\end{equation}}
 \newcommand{\bea}{\begin{eqnarray}}
 \newcommand{\eea}{\end{eqnarray}}
 \def\disp{\displaystyle}

%
 \newcommand{\dmunit}{ {\rm pc\hspace{0.24em} cm^{-3}} }
 \newcommand{\cmn}{,\,}

 \begin{titlepage}

 \begin{flushright}
 arXiv:2203.12551
 \end{flushright}

 \title{\Large \bf A Possible Subclassification of Fast Radio Bursts}

 \author{Han-Yue~Guo\,}
 \email[\,email address:\ ]{guohanyue7@163.com}
 \affiliation{School of Physics,
 Beijing Institute of Technology, Beijing 100081, China}

 \author{Hao~Wei\,}
 \email[\,Corresponding author;\ email address:\ ]{haowei@bit.edu.cn}
 \affiliation{School of Physics,
 Beijing Institute of Technology, Beijing 100081, China}

 \begin{abstract}\vspace{1cm}
 \centerline{\bf ABSTRACT}\vspace{2mm}
 Although fast radio bursts (FRBs) have been an active field in
 astronomy and cosmology, their origin is still unknown to date. One of the
 interesting topics is the classification of FRBs, which is closely related
 to the origin of FRBs. Different physical mechanisms are required by
 different classes of FRBs. In the literature, they usually could be
 classified into non-repeating and repeating FRBs. Well motivated by the
 observations, here we are interested in the possible subclassification of
 FRBs. By using the first CHIME/FRB catalog, we propose to subclassify
 non-repeating (type~I) FRBs into type~Ia and Ib FRBs. The distribution
 of type~Ia FRBs is delayed with respect to the cosmic star formation
 history (SFH), and hence they are probably associated with old stellar
 populations, while the distribution of type~Ib FRBs tracks SFH, and hence
 they are probably associated with young stellar populations. Accordingly,
 the physical criteria for this subclassification of type~I FRBs have
 been clearly determined. We find that there are some tight empirical
 correlations for type~Ia FRBs but not for type~Ib FRBs, and vice versa.
 These make them different in physical properties. Similarly, we suggest
 that repeating (type~II) FRBs could also be subclassified into type~IIa
 and IIb FRBs. A universal subclassification scheme is given at the end.
 This subclassification of FRBs might help us to reveal quite
 different physical mechanisms behind them, and improve
 their applications in astronomy and cosmology.
 \end{abstract}


 \maketitle

 \end{titlepage}

 \renewcommand{\baselinestretch}{1.0}


\section{Introduction}\label{sec1}

Although fast radio bursts (FRBs) have been an active field in astronomy
 and cosmology, their origin is still unknown to date. FRBs are
 mysterious transient radio sources of millisecond duration~\cite{NAFRBs,
 Lorimer:2018rwi,Keane:2018jqo,Caleb:2018ygr,Pen:2018ilo,Petroff:2019tty,
 Petroff:2021wug,Zhang:2020qgp,Lyubarsky:2021bai,Xiao:2021omr,Caleb:2021xqe,
 Nicastro:2021cxs,Cordes:2019cmq,Pilia:2022mul,Bhandari:2021thi}. Most of
 them are at extragalactic/cosmological distances, as suggested by their
 large dispersion measures~(DMs) well in excess of the Galactic values,
 and hence FRBs are a promising probe to study cosmology and
 the intergalactic medium (see e.g.~\cite{Qiang:2021ljr,Deng:2013aga,
 Yang:2016zbm,Gao:2014iva,Zhou:2014yta,Qiang:2019zrs,Qiang:2020vta,
 Qiang:2021bwb} and also \cite{NAFRBs,Lorimer:2018rwi,Keane:2018jqo,
 Caleb:2018ygr,Pen:2018ilo,Petroff:2019tty,Petroff:2021wug,Zhang:2020qgp,
 Lyubarsky:2021bai,Xiao:2021omr,Caleb:2021xqe,Nicastro:2021cxs,
 Cordes:2019cmq,Pilia:2022mul,Bhandari:2021thi}).

To reveal the possible origins of FRBs, various topics are extensively
 debated in the literature~\cite{NAFRBs,Lorimer:2018rwi,Keane:2018jqo,
 Caleb:2018ygr,Pen:2018ilo,Petroff:2019tty,Petroff:2021wug,Zhang:2020qgp,
 Lyubarsky:2021bai,Xiao:2021omr,Caleb:2021xqe,Nicastro:2021cxs,
 Cordes:2019cmq,Pilia:2022mul,Bhandari:2021thi}, such as the engine,
 radiation mechanism, distribution, classification, propagation effect,
 and cosmological application of FRBs. Many theoretical models have
 been proposed to this end, and we refer to e.g.~\cite{Petroff:2019tty,
 Petroff:2021wug,Zhang:2020qgp,Lyubarsky:2021bai,Xiao:2021omr,Caleb:2021xqe,
 Nicastro:2021cxs,Cordes:2019cmq} for comprehensive reviews and
 \cite{Platts:2018hiy} for the up-to-date online catalogue of FRB theories.
 On the other hand, the observational data were rapidly accumulated in
 the recent years~\cite{Xiao:2021omr,Caleb:2021xqe,Nicastro:2021cxs,
 Pilia:2022mul,Cordes:2019cmq,Petroff:2016tcr,Heintz:2020}. Therefore, many
 impressive progresses have been made in the field of FRBs.

One of the interesting topics is the classification of FRBs
 \cite{Caleb:2018ygr,Petroff:2019tty,Petroff:2021wug,Zhang:2020qgp,
 Xiao:2021omr}. How many different populations of FRBs exsist? In the actual
 observations, many FRBs were found to be (apparently) one-off, while
 some FRBs are repeating. So, it is natural to classify them into two
 populations: non-repeating FRBs and repeating FRBs. This classification is
 closely related to the origin of FRBs. Obviously, the repeaters rule out
 the cataclysmic engines for these sources. However, the question is
 whether the apparently non-repeating FRBs are genuinely one-off or not. In
 fact, some apparently non-repeating FRBs were found to be repeaters in the
 follow-up observations. It is possible that all FRBs repeat, and the
 non-detection of repetition might be due to the long waiting time or
 low flux of the repeating bursts (see e.g.~\cite{Palaniswamy:2017aze,
 Ai:2020wnm,Caleb:2019szc}), or an unknown selection effect
 \cite{Connor:2020phs}. Some unified models for repeating and non-repeating
 FRBs were proposed in the literature (see e.g.~\cite{Bagchi:2017tzi,
 Yamasaki:2017hdr,Katz:2022cyt,Katz:2022sqt}). Recently, some works have
 tried to address this question. In~\cite{Ai:2020wnm}, the number fraction
 of repeating FRBs was predicted to peak at a value less than $100\%$ in the
 future if non-repeating FRBs are genuinely one-off, otherwise it will
 increase to $100\%$ eventually. In~\cite{Hashimoto:2020acj}, it was found
 that the time-integrated-luminosity functions and volumetric occurrence
 rates of non-repeating and repeating FRBs against redshift are
 significantly different. In~\cite{Zhong:2022uvu}, it was claimed that the
 discriminant properties in FRBs is difficult to be explained by a single
 population. In~\cite{Pleunis:2021qow}, an observed difference in the burst
 morphologies of one-off FRBs and repeater bursts was found. The above
 works indicate that it is reasonable to classify them into repeating and
 non-repeating FRBs.

Another natural question is whether there are other classifications of FRBs
 different from repeating and non-repeating FRBs. Recently, several efforts
 were made in the literature. In~\cite{Li:2021yds}, similar to
 gamma-ray bursts (GRBs), it was proposed to classify FRBs into short
 ($<100\,\rm ms$) and long ($>100\,\rm ms$) FRBs. A tight power-law
 correlation between fluence and peak flux density was found for them. Long
 FRBs are more energetic than short FRBs in the fluence versus extragalactic
 DM plane. In~\cite{Xiao:2021viy}, it was argued that the brightness
 temperature $T_B$ might be used to classify the repeating bursts into
 classical ($T_B\geq 10^{33}\,\rm K$) and atypical ($T_B<10^{33}\,\rm K$)
 ones in the light of the well-known repeating FRB 20121102A. A tight
 power-law correlation between pulse width and fluence was found for
 classical bursts. In~\cite{Chaikova:2022vnh}, using cross-correlation and
 clustering algorithms applied to one-dimensional intensity profiles of the
 bursts, two major classes of FRBs featuring different waveform morphologies
 and simultaneously different distributions of brightness temperature were
 identified. These efforts might shed new light on the nature of FRBs.

In the present work, we are interested in the possible subclassification of
 FRBs. There are two main motivations for doing this. The first one comes
 from the neighboring fields of supernovae and GRBs. As is well known, there
 are two major classes of supernovae: type I and II~\cite{SN}. Then, type I
 supernovae are subclassified into type Ia, Ib and Ic, while type II
 supernovae are subclassified into type II-P, II-L, IIn and IIb. Only the
 well-known type Ia supernovae could be used as standard candles, which led
 to the great discovery of cosmic acceleration (and Nobel prize in physics
 2011). This highlights the importance of the subclassification. On the
 other hand, GRBs are usually classified into long and short ones. However,
 the existence of temporally long events showing signatures of short GRBs
 led to introduce an alternative classification: type I (typically short
 and associated with old populations) and type II (typically long and
 associated with young populations)~\cite{Zhang:2006mb,Kumar:2014upa}.
 Similarly, our second motivation is related to FRBs associated with young
 or old populations. For a long time, it was speculated that the FRB
 distribution tracks the cosmic star formation history (SFH)~\cite{NAFRBs,
 Lorimer:2018rwi,Keane:2018jqo,Caleb:2018ygr,Pen:2018ilo,Petroff:2019tty,
 Petroff:2021wug,Zhang:2020qgp,Lyubarsky:2021bai,Xiao:2021omr,Caleb:2021xqe,
 Nicastro:2021cxs,Cordes:2019cmq,Pilia:2022mul,Bhandari:2021thi}. The
 landmark Galactic FRB 200428 associated with the young magnetar SGR
 1935+2154 \cite{Andersen:2020hvz,Bochenek:2020zxn,Lin:2020mpw,Li:2020qak}
 confirmed that at least some (if not all) FRBs originate from young
 magnetars. So, it is reasonable to expect that the FRB distribution is
 closely correlated with star-forming activities, as observed for the
 repeating FRB 121102~\cite{Tendulkar:2017vuq}, FRB 180916.J0158+65
 \cite{Marcote:2020ljw}, FRB 20190520B~\cite{Niu:2021bnl}, FRB 20181030A
 \cite{Bhardwaj:2021hgc}, and FRB 20201124A~\cite{Piro:2021upe,
 Nimmo:2021ntn}. But it was argued in~\cite{Xu:2021qdn} that FRB 20201124A
 is located at an inter-arm region of a barred-spiral galaxy, namely an
 environment not directly expected for young populations. On the other hand,
 the recently discovered repeating FRB 20200120E in a globular cluster of
 the nearby galaxy M81 \cite{Bhardwaj:2021xaa,Kirsten:2021llv,Nimmo:2021yob}
 suggested that some FRBs are associated with old stellar populations.
 In~\cite{Zhang:2021kdu}, it was claimed that the bursts of the first
 CHIME/FRB catalog~\cite{CHIMEFRB:2021srp} as a whole do not track SFH.
 In~\cite{Qiang:2021ljr}, it was independently confirmed that the FRB
 distribution model tracking SFH can be rejected at high confidence, and
 a suppressed evolution (delay) with respect to SFH was found. Putting the
 above facts together, it is reasonable to speculate that some FRBs are
 associated with young populations and hence they track SFH, while the other
 FRBs are associated with old populations and hence they do not track
 SFH. So, a possible subclassification of FRBs is required.

This paper is organized as followings. In Sec.~\ref{sec2}, we briefly
 introduce the observational data of FRBs, namely the first CHIME/FRB
 catalog~\cite{CHIMEFRB:2021srp}. In Sec.~\ref{sec3}, we show that the
 distributions of non-repeating and repeating FRBs are significantly
 different. For convenience, we suggest calling them type I and II FRBs,
 respectively. Then, we propose to subclassify non-repeating (type~I) FRBs
 into type Ia and Ib FRBs. The distribution of type Ia FRBs is delayed with
 respect to SFH, and hence they are probably associated with old stellar
 populations, while the distribution of type Ib FRBs tracks SFH, and hence
 they are probably associated with young stellar populations. Accordingly,
 the physical criteria for this subclassification have been clearly
 determined. In Sec.~\ref{sec4}, we find that there are some tight empirical
 correlations for type Ia FRBs but not for type Ib FRBs, and vice versa.
 These empirical correlations make them different in physical properties.
 Clearly, type Ia and Ib FRBs require quite different physical mechanisms.
 In Sec.~\ref{sec5}, we turn to repeating (type~II) FRBs. In
 Sec.~\ref{sec6}, some brief concluding remarks and a universal
 subclassification scheme are given.


\section{The observational data}\label{sec2}

As is well known, one of the key observational quantities of
 FRBs is the dispersion measure DM, namely the column density of the
 free electrons, due to the ionized medium (plasma) along the
 path. Clearly, the observed DM of FRB can be separated into
 \cite{Qiang:2021ljr,Deng:2013aga,Yang:2016zbm,Gao:2014iva,
 Zhou:2014yta,Qiang:2019zrs,Qiang:2020vta,Qiang:2021bwb}
 \be{eq1}
 {\rm DM_{obs}=DM_{MW}+DM_{halo}+DM_{IGM}+DM_{host}}/(1+z)\,,
 \ee
 where $z$ is the redshift, and $\rm DM_{MW}$, $\rm DM_{halo}$,
 $\rm DM_{IGM}$, $\rm DM_{host}$ are the contributions from the
 Milky Way, the Milky Way halo, the intergalactic medium (IGM), the
 host galaxy (including interstellar medium of the host galaxy
 and the near-source plasma), respectively. For convenience,
 one could introduce the extragalactic DM~\cite{Qiang:2021ljr,Deng:2013aga,
 Yang:2016zbm,Gao:2014iva,Zhou:2014yta,Qiang:2019zrs,Qiang:2020vta,
 Qiang:2021bwb}, namely
 \be{eq2}
 {\rm DM_E=DM_{obs}-DM_{MW}-DM_{halo}=DM_{IGM}+DM_{host}}/(1+z)\,.
 \ee
 Here, we adopt $\rm DM_{halo}=30\,\dmunit$ (see e.g.~\cite{Dolag:2014bca,
 Prochaska:2019mn}), and $\rm DM_{host}=50\,\dmunit$ (see
 e.g.~\cite{Shannon:2018,Prochaska:2019sci,Hashimoto:2019aqu}).
 In fact, they are the ones used in the literature for our Milky Way at
 high Galactic latitude. We can obtain $\rm DM_{MW}$ by using
 NE2001~\cite{Cordes:2002wz,Cordes:2003ik,pygedm} up to $30\,\rm kpc$.
 $\rm DM_{IGM}$ is given by~\cite{Qiang:2021ljr,Deng:2013aga,Yang:2016zbm,
 Gao:2014iva,Zhou:2014yta,Qiang:2019zrs,Qiang:2020vta,Qiang:2021bwb}
 \be{eq3}
 {\rm DM_{IGM}}=\frac{3cH_0\Omega_{b}}{8\pi G m_p}
 \int_0^z\frac{f_{\rm IGM}(\tilde{z})\,f_e(\tilde{z})\left(1+
 \tilde{z}\right)d\tilde{z}}{h(\tilde{z})}\,,
 \ee
 where $c$ is the speed of light, $H_0$ is the Hubble constant, $\Omega_b$
 is the present fractional density of baryons, $G$ is the gravitational
 constant, $m_p$ is the mass of proton, $h(z)\equiv H(z)/H_0$ is the
 dimensionless Hubble parameter, $f_{\rm IGM}(z)$ is the fraction of baryon
 mass in IGM, and $f_e(z)$ is the ionized electron number fraction per
 baryon. The latter two are functions of redshift $z$ in principle.
 Following e.g.~\cite{Hashimoto:2020acj,Zhou:2014yta,Zhang:2021kdu,
 Qiang:2021ljr}, we use the fiducial values $f_e=7/8$ and $f_{\rm IGM}=
 0.82$ in this work.~Note that it is very safe to adopt $f_e=(3/4)\,
 \chi_{e\cmn\rm H}(z)+(1/4)\,\chi_{e\cmn\rm He}(z)=7/8$ for
 $\chi_{e\cmn\rm H}(z)=\chi_{e\cmn\rm He}(z)=1$, since hydrogen and
 helium are both fully ionized at $z\leq 3$ for almost all observed FRBs
 \cite{Qiang:2021ljr,Deng:2013aga,Yang:2016zbm,Gao:2014iva,Zhou:2014yta,
 Qiang:2019zrs,Qiang:2020vta,Qiang:2021bwb}. Actually, the variation of
 $f_{\rm IGM}(z)$ is fairly small as it could be constrained by using other
 cosmological observations such as cosmic microwave background (see
 e.g.~\cite{Deng:2013aga}), and hence it is also reasonable to adopt a
 constant $f_{\rm IGM}$ at low redshifts $z\leq 3$. On the other hand,
 we consider the fiducial cosmology in this work, namely the well-known
 flat $\Lambda$CDM model, and hence
 \be{eq4}
 h(z)=\left[\Omega_m\left(1+z\right)^3+\left(1-\Omega_m
 \right)\right]^{1/2}\,,\quad\quad
 d_L=\left(1+z\right)d_C=c\left(1+z\right)
 \int_0^z\frac{d\tilde{z}}{H(\tilde{z})}\,,
 \ee
 where $d_L$ and $d_C$ are the luminosity distance and the comoving
 distance, respectively. In this work, we adopt
 $\Omega_m=0.3153$, $\Omega_b=0.0493$, and $H_0=67.36\,{\rm km/s/Mpc}$ from
 the Planck 2018 results~\cite{Aghanim:2018eyx}.

The first CHIME/FRB catalog~\cite{CHIMEFRB:2021srp} of 536 events (including
 474 one-off bursts and 62 repeat bursts from 18 repeaters) was released in
 June 2021. Such a large uniform sample detected by a single telescope is
 very valuable to study FRBs. We preliminarily deal with it following
 \cite{Qiang:2021ljr}. At first, we exclude the bursts with zero fluences
 and the bursts labeled with $\tt excluded_{-}flag=1$. Following e.g.
 \cite{Zhang:2021kdu,Qiang:2021ljr}, we only use the first detected burst of
 each FRB source. In practice, we identify the non-repeaters labeled with
 $\tt repeater_{-}name=-9999$ and then only take the ones labeled with
 $\tt sub_{-}num=0$ (434 bursts in total). We identify the repeaters labeled
 with $\tt repeater_{-}name\not=-9999$ and only take the ones labeled with
 $\tt sub_{-}num=0$, and then from them we adopt the first ones in each
 group with the same $\tt repeater_{-}name$ (18 bursts in total). For each
 burst, its ``\,observed\,'' $\rm DM_E=DM_{obs}-DM_{MW}-DM_{halo}$,
 where $\rm DM_{obs}$ is given by the column labeled with
 ``\,$\tt bonsai_{-}dm$\,'' in the data table. Then, its inferred redshift
 $z$ is obtained by numerically solving ${\rm DM_E=DM_{IGM}+
 DM_{host}}/(1+z)$ with $\rm DM_{IGM}$ given by Eq.~(\ref{eq3}). In
 this work, we require a very conservative criterion $\rm DM_{IGM}\geq
 DM_{obs}/10$ to exclude the bursts very close to us. After this robust cut,
 we have 430 one-off FRBs and 17 repeaters.

For each burst, its observed specific fluence $F_\nu$ is given by the column
 labeled with ``\,{\tt fluence}\,'' in the data table. Assuming a flat radio
 spectrum, the specific fluence is related with isotropic energy $E$
 according to~\cite{Zhang:2018csb,Zhang:2020ass,Zhang:2021kdu,Qiang:2021ljr}
 \be{eq5}
 F_\nu=\frac{\left(1+z\right)E}{4\pi d_L^2 \nu_c}\,,
 \ee
 where $\nu_c$ is the central observing frequency. For CHIME, $\nu_c=600\,
 {\rm MHz}$~\cite{CHIMEFRB:2021srp}. Thus, the ``\,observed\,'' isotropic
 energy $E$ can be inferred from Eq.~(\ref{eq5}) with the observed $F_\nu$
 and the luminosity distance $d_L$ given by Eq.~(\ref{eq4}). So far, the
 observational data of 430 one-off FRBs and 17 repeaters are ready.


\section{Subclassification of non-repeating FRBs}\label{sec3}


\subsection{Type I and II FRBs}\label{sec3a}

In the radio sky, there are many known transients besides FRBs, such
 as pulsars, solar bursts, rotating radio transients (RRATs),
 nano-shots, flare stars/brown dwarves, X-ray binaries, RSCVn/Algols,
 novae, supernovae, AGN/blazar/QSO, giant radio pulses (GRPs), and GRBs.
 We refer to e.g. Fig.~1 of~\cite{Keane:2018jqo} for details. As is well
 known, they could be well distinguished in the transient duration $\nu W$
 versus spectral luminosity $L_\nu$ phase plane, with the help of brightness
 temperature $T_B$ which relates to the radiation mechanism
 (see e.g. Fig.~1 of~\cite{Keane:2018jqo}, Fig.~5 of~\cite{Pietka:2014wra},
 Fig.~3 of~\cite{Nimmo:2021yob}, Fig.~7 of~\cite{Petroff:2021wug}, and
 Fig.~4 of~\cite{Majid:2021uli}).

\begin{center}
 \begin{figure}[tb]
 \centering
 \vspace{-10mm}  
 \includegraphics[width=0.75\textwidth]{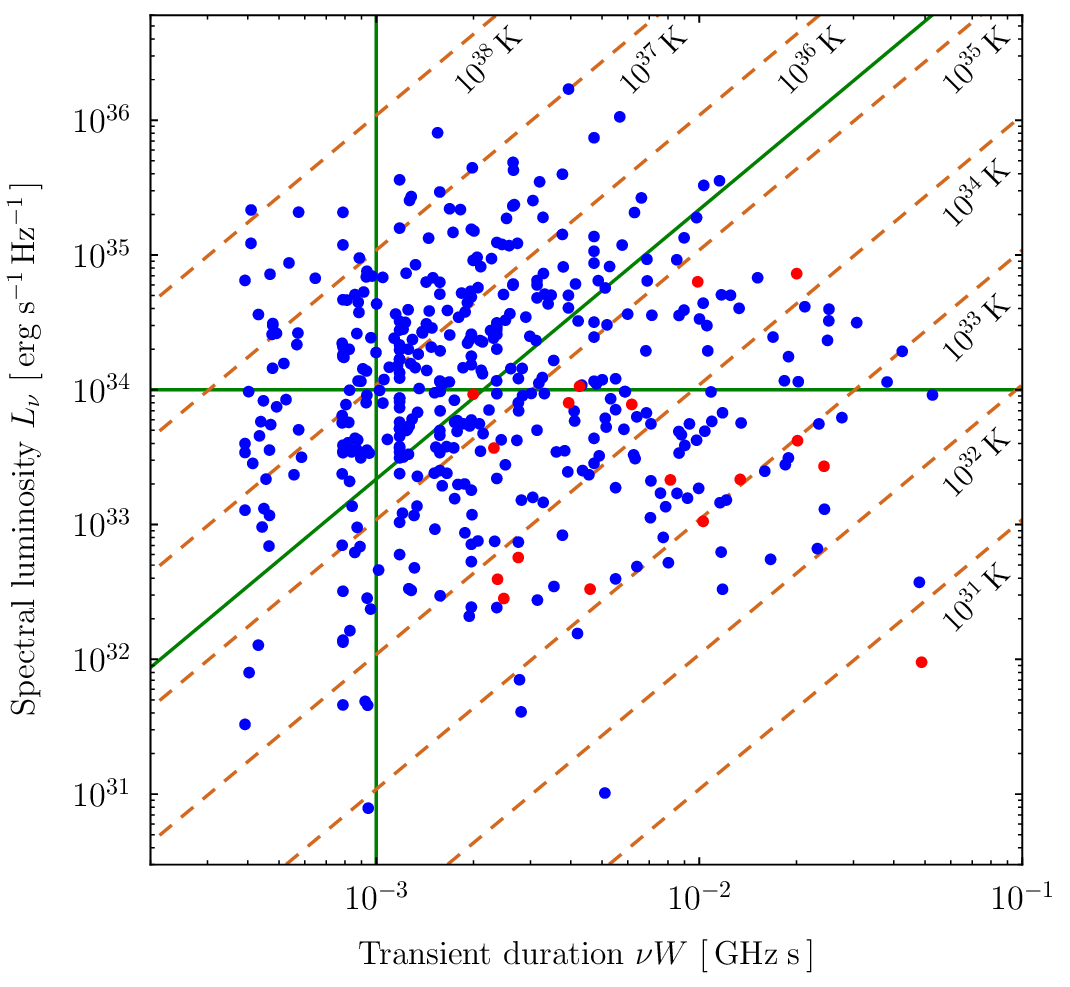}
 \caption{\label{fig1} 430 one-off FRBs (blue points) and 17 repeaters
 (red points) from the first CHIME/FRB catalog in the transient duration
 $\nu W$ versus spectral luminosity $L_\nu$ plane, with some isothermal
 lines of brightness temperature $T_B$ (chocolate dashed lines). The green
 solid lines $\nu W=10^{-3}\,{\rm GHz\;s}$, $L_\nu=10^{34}\,{\rm erg/s/Hz}$
 and $T_B=2\times 10^{35}\,{\rm K}$ are used to divide this $\nu W-L_\nu$
 phase plane into various regions. See Sec.~\ref{sec3} and
 Fig.~\ref{fig3} for details.}
 \end{figure}
 \end{center}

Since the $\nu W-L_\nu$ phase plane is very useful to distinguish radio
 transients, we might also use it to subclassify FRBs. In Fig.~\ref{fig1},
 we plot 430 one-off FRBs and 17 repeaters from the CHIME/FRB catalog in the
 $\nu W-L_\nu$ plane, with some isothermal lines of $T_B$. For each burst,
 we take the frequency $\nu$ and the pulse width $W$ from the columns
 labeled with ``\,{\tt peak$_{-}$freq}\,'' and ``\,{\tt bc$_{-}$width}\,''
 in the CHIME/FRB data table, respectively. We calculate the spectral
 luminosity $L_\nu$ according to (e.g.~\cite{Pietka:2014wra,Majid:2021uli})
 \be{eq6}
 L_\nu=4\pi d_L^2 S_\nu\,,
 \ee
 where the luminosity distance $d_L$ is given by Eq.~(\ref{eq4}), and the
 flux $S_\nu$ is given by the column labeled with ``\,{\tt flux}\,'' in
 the CHIME/FRB data table. The brightness temperature $T_B$ is given by
 (e.g.~\cite{Xiao:2021viy,Pietka:2014wra,Majid:2021uli})
 \be{eq7}
 T_B=\frac{S_\nu\, d_L^2}{2\pi\kappa_B\left(\nu W\right)^2}
 =1.1\times 10^{35}\,{\rm K}\,\left(\frac{S_\nu}{\rm Jy}\right)
 \left(\frac{d_L}{\rm Gpc}\right)^2\left(\frac{\nu}{\rm
 GHz}\right)^{-2}\left(\frac{W}{\rm ms}\right)^{-2}\,,
 \ee
 which can be expressed in terms of $\nu W$ and $L_\nu$\,. $\kappa_B$ is
 the Boltzmann constant. It is worth noting that the above quantities might
 be slightly different in the literature (for example, in some works,
 the angular diameter distance $d_A$ might be used instead of $d_L$, the
 central frequency $\nu_c$ might be used instead of the peak frequency,
 $\pi$ might be removed, the redshift might be introduced, and so on).
 We intentionally use them as the ones mentioned above, because they
 work well for our purpose in the present forms.

\begin{center}
 \begin{figure}[tb]
 \centering
 \vspace{-10mm}  
 \includegraphics[width=0.9\textwidth]{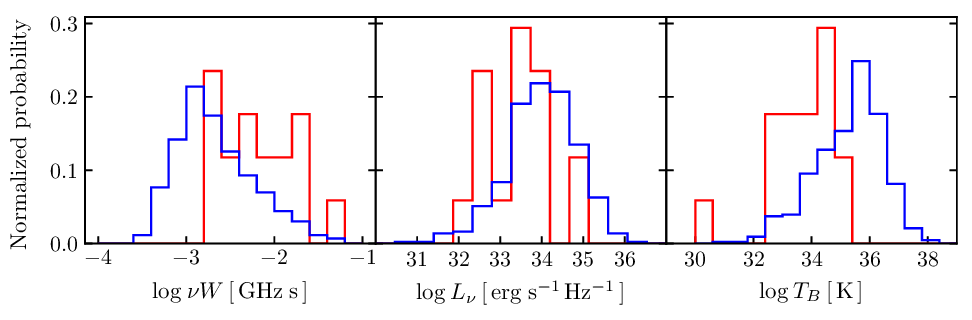}
 \caption{\label{fig2} Normalized $\nu W$, $L_\nu$ and $T_B$ distributions
 of 430 non-repeaters (blue histograms) and 17 repeaters (red histograms),
 respectively. See Sec.~\ref{sec3a} for details.}
 \end{figure}
 \end{center}

From Fig.~\ref{fig1}, it is easy to see that the distributions
 of non-repeating and repeating FRBs are different in the $\nu W-L_\nu$
 phase plane. Clearly, most of the repeaters are located in the
 bottom-right region, where the transient duration $\nu W$ is relatively
 large, the spectral luminosity $L_\nu$ is relatively low, and
 the brightness temperature $T_B$ is also relatively low. In
 Fig.~\ref{fig2}, we present the normalized $\nu W$, $L_\nu$ and $T_B$
 distributions of non-repeating and repeating FRBs, respectively. Again, we
 find that these distributions are clearly different for non-repeaters and
 repeaters (note that the number of repeaters is only 17, and hence they do
 not form a good enough statistics). Thus, its is reasonable to classify
 FRBs into two types as usual. For convenience, we suggest
 calling non-repeating/repeating FRBs type I/II FRBs, respectively.


\vspace{-12mm} 


\subsection{Type Ia and Ib FRBs}\label{sec3b}

At first, we consider the possible subclassification of non-repeating
 (type~I) FRBs. By definition, the subclasses should be significantly
 different. As is well known, the Kolmogorov-Smirnov (KS) test is one of the
 useful tools to compare two samples~\cite{KStest}. One can perform the KS
 test by using {\tt scipy.stats.kstest} in Python~\cite{KStestpy}, which
 returns the KS statistic and the corresponding p-value. For convenience, we
 use the p-value ($p_{_{\rm KS}}$) in the two-sample case as in
 \cite{CHIMEFRB:2021srp,Chawla:2021igg,Qiang:2021ljr}, rather than the KS
 statistic ($D_{\rm KS}$). The null hypothesis (namely two samples are drawn
 from the same distribution) can be rejected at $90\%$ ($95\%$) confidence
 if $p_{_{\rm KS}}<0.1$ ($0.05$), respectively. Otherwise, two samples can
 be consistent with each other if $p_{_{\rm KS}}>0.1$ (or $0.05$). $p_{_{\rm
 KS}}$ is higher for two closer samples (and $p_{_{\rm KS}}=1$ for two
 identical samples), while $p_{_{\rm KS}}\to 0$ for two completely different
 samples.


 \begin{center}
 \begin{figure}[tb]
 \centering
 \vspace{0.5mm}  
 \includegraphics[width=0.75\textwidth]{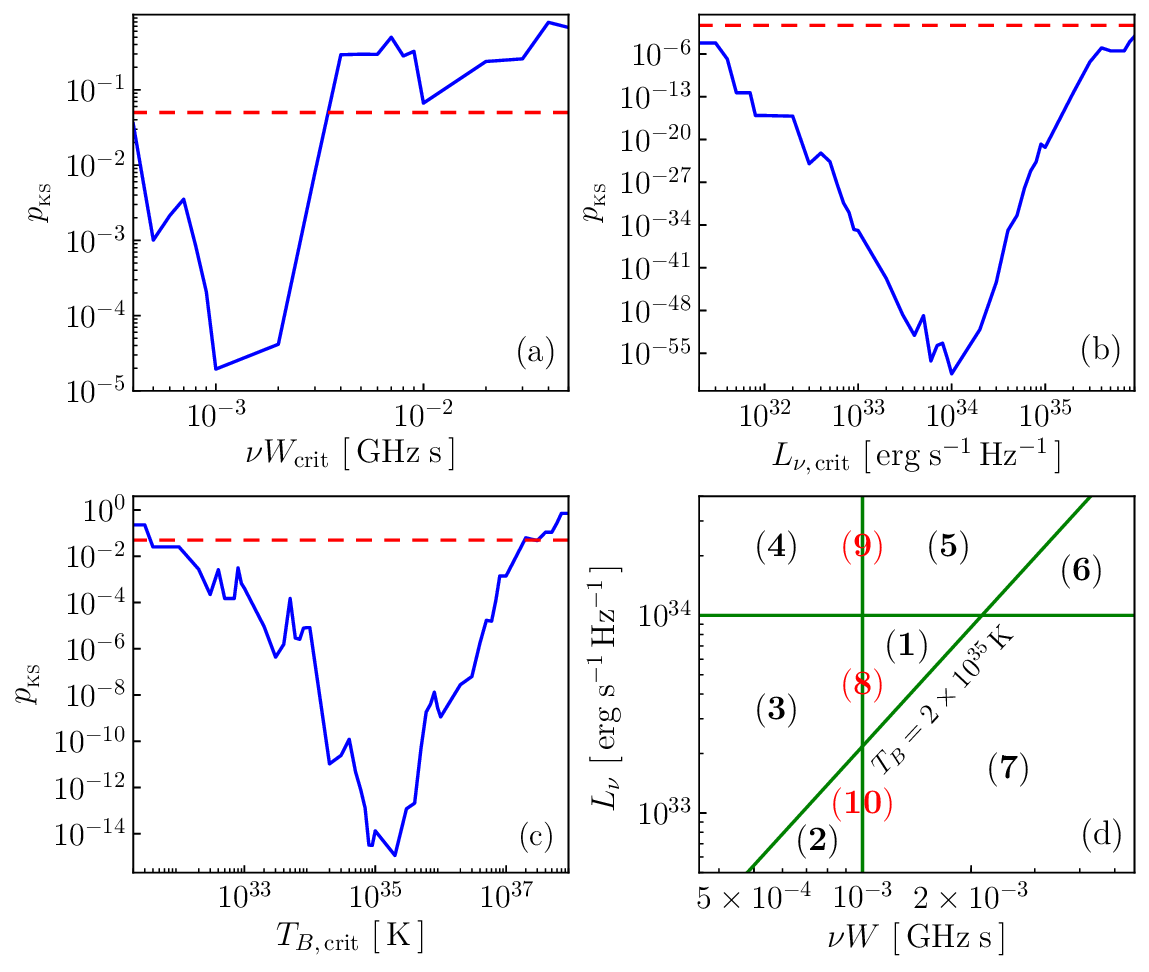}
 \caption{\label{fig3} The p-values $p_{_{\rm KS}}$ are shown as functions
 of $\nu W_{\rm crit}$ (a), $L_{\nu\cmn\rm crit}$ (b) and $T_{B\cmn\rm
 crit}$ (c). $p_{_{\rm KS}}=0.05$ is indicated by the red dashed lines. In
 panel (d), the $\nu W-L_\nu$ phase plane is divided into seven regions by
 the green solid lines $\nu W=10^{-3}\,{\rm GHz\;s}$, $L_\nu=10^{34}\,
 {\rm erg/s/Hz}$ and $T_B=2\times 10^{35}\,{\rm K}$. These 7 regions are
 labeled with the numbers $(1)\sim (7)$. In addition, region~(8) = regions
 $(1)+(3)$, region~(9) = regions $(4)+(5)$, region~(10) = regions $(2)+(7)$.
 Panel (d) should be viewed together with Fig.~\ref{fig1}. See
 Sec.~\ref{sec3b} for details.}
 \end{figure}
 \end{center}



 \begin{center}
 \begin{figure}[tb]
 \centering
 \vspace{-10mm}  
 \includegraphics[width=0.83\textwidth]{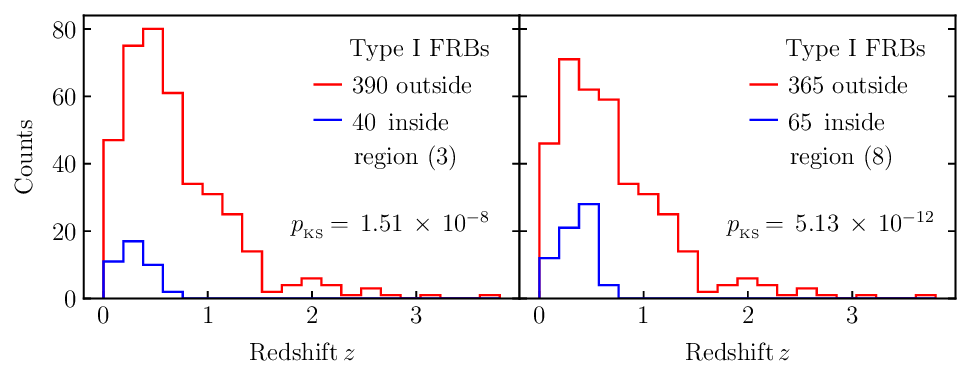}
 \caption{\label{fig4} Left panel: The redshift distributions of type~I
 FRBs inside (blue) and outside (red) region~(3). The numbers of these two
 samples are given, respectively. We also present the p-value $p_{_{\rm
 KS}}$ of the KS test for these two redshift distributions. Right panel:
 The same as in left panel, but for region~(8). See Sec.~\ref{sec3b} for
 details.}
 \end{figure}
 \end{center}


\vspace{-25.8mm} 

There are three candidates ($\nu W$, $L_\nu$ and $T_B$) in the $\nu W-L_\nu$
 phase plane for the possible criteria of subclassification. For a given
 critical value (say, $\nu W_{\rm crit}$), 430 non-repeating FRBs can be
 divided into two samples (say, $\nu W\geq\nu W_{\rm crit}$ and $\nu W<\nu
 W_{\rm crit}$), and then we compare the redshift distributions of these two
 samples by using the KS test. So, the p-values $p_{_{\rm KS}}$ can be found
 as functions of $\nu W_{\rm crit}$, $L_{\nu\cmn\rm crit}$ and $T_{B\cmn\rm
 crit}$ by scanning their parameter spaces, respectively. We show them in
 Fig.~\ref{fig3}. Note that two samples having a lower $p_{_{\rm KS}}$ are
 more different. Thus, the minimal p-values indicate the best dividing
 lines. From panels (a), (b), (c) of Fig.~\ref{fig3}, we find that the best
 dividing lines are given by $\nu W=10^{-3}\,{\rm GHz\;s}$, $L_\nu=10^{34}\,
 {\rm erg/s/Hz}$ and $T_B=2\times 10^{35}\,{\rm K}$. They are plotted as the
 green solid lines in Fig.~\ref{fig1} and panel (d) of Fig.~\ref{fig3}.
 Clearly, they divide the $\nu W-L_\nu$ phase plane into seven regions,
 as labeled by the numbers $(1)\sim (7)$ in panel (d) of Fig.~\ref{fig3}.

As mentioned above, we speculate that some FRBs track SFH and the others
 do not. Their distributions should be significantly different. So far, we
 have divided the $\nu W-L_\nu$ phase plane into 7 regions as above. For a
 given region, 430 non-repeating (type~I) FRBs inside/outside this region
 form two samples. One can compare them by using the KS test. But now it is
 more important to see whether one of these two samples tracks SFH.
 Fortunately, a suitable method for this purpose was proposed
 in~\cite{Zhang:2021kdu} and then has been extended in~\cite{Qiang:2021ljr}.
 Here, we closely follow the method used in~\cite{Qiang:2021ljr}. The key
 idea is to confront the Monte Carlo simulations with the observational
 data. If the simulations are rejected by the observational data, the
 assumed FRB distribution models generating these simulations could be ruled
 out. Otherwise, they survive. In the present work, we generate
 the simulations assuming SFH, namely the mock observed FRB redshift rate
 distribution is given by~\cite{Zhang:2021kdu,Qiang:2021ljr}
 \be{eq8}
 \frac{dN}{dt_{\rm obs}\,dz}=\frac{1}{1+z}\cdot\frac{dN}{dt\,dV}\cdot
 \frac{dV}{dz}=\frac{1}{1+z}\cdot\frac{dN}{dt\,dV}\cdot\frac{c}{H_0}
 \cdot\frac{4\pi d_C^2}{h(z)}\,,
 \ee
 where we have used $dt/dt_{\rm obs}=(1+z)^{-1}$ due to the
 cosmic expansion, $d_C=d_L/(1+z)$ and $h(z)$ are given by Eq.~(\ref{eq4}),
 and we assume that the distribution of mock FRBs track
 SFH~\cite{Madau:2016jbv,Qiang:2021ljr}, namely
 \be{eq9}
 \frac{dN}{dt\,dV}\propto {\rm SFH}(z)\propto\frac{(1+z)^{2.6}}
 {1+\left((1+z)/3.2\right)^{6.2}}\,.
 \ee
 Note that Eq.~(\ref{eq9}) is the best-fit SFH density from the latest
 observational data of ultraviolet and infrared surveys (see Eq.~(1)
 of~\cite{Madau:2016jbv}), which characterizes the real SFH of our universe.
 On the other hand, we generate the isotropic energy $E$ for the mock
 FRBs with~\cite{Zhang:2021kdu,Qiang:2021ljr}
 \be{eq10}
 dN/dE\propto\left(E/E_c\right)^{-\alpha}\exp\left(-E/E_c\right)\,,
 \ee
 where $\alpha=1.9$ and $\log\left(E_c/{\rm erg}\right)=41$ are fixed as
 in~\cite{Qiang:2021ljr}, while ``\,$\log$\,'' gives the logarithm to base
 10. Notice that Eq.~(\ref{eq10}) corresponds to the Schechter luminosity
 function of FRBs~\cite{Luo:2020wfx,Schechter:1976iz}.~Actually, the
 isotropic energy distribution in Eq.~(\ref{eq10}) is characterized by a
 simple power law $\propto E^{-\alpha}$ with a sharp exponential cutoff
 around the energy scale $E_c\,$. In the literature, $\alpha$ and $E_c$
 could be constrained by the observations, i.e.~$1.8\lesssim\alpha\lesssim
 2$ roughly~\cite{Zhang:2020ass,Luo:2020wfx,Lu:2019pdn,Luo:2018tiy}
 and $E_c\sim 3\times 10^{41}\,{\rm erg}$ loosely~\cite{Luo:2020wfx,
 Zhang:2020ass}. So, $\alpha=1.9$ and $\log\left(E_c/{\rm erg}\right)=41$
 are well consistent with the observations.~One can generate $N_{\rm sim}$
 mock FRBs tracking SFH as follows: (i) randomly assign a mock redshift
 $z_i$ to the $i$-th mock FRB from the redshift distribution in
 Eq.~(\ref{eq8}) with Eq.~(\ref{eq9}); (ii) generate a mock energy $E_i$
 randomly from the distribution in Eq.~(\ref{eq10}) for this mock FRB;
 (iii) derive the specific fluence $F_{\nu\cmn i}$ by using Eq.~(\ref{eq5})
 with $z_i$ and $E_i$ for this mock FRB; (iv) derive ${\rm DM}_{{\rm E}
 \cmn i}$ at redshift $z_i$ by using Eq.~(\ref{eq2}) with Eq.~(\ref{eq3})
 for this mock FRB; (v) repeat the above steps for $N_{\rm sim}$ times.
 Finally, $N_{\rm sim}$ mock FRBs are on hand.


\begin{table}[tb]
 \renewcommand{\arraystretch}{1.7}
 \begin{center}
 \vspace{-8mm}  
 \begin{tabular}{cccc} \hline\hline
 \quad $\log F_{\nu\cmn {\rm th}}^{\rm max}$ \quad
 & \quad $p_{_{\rm KS}}$ for $\log F_\nu$ \quad
 & \quad $p_{_{\rm KS}}$ for $\log E$ \quad
 & \quad $p_{_{\rm KS}}$ for $\rm DM_E$ \quad \\ \hline
 {\bf 0.77} & {\bf 0.2750} & {\bf 0.2859} & {\bf 0.4575} \\ \hline
 0.78 & 0.3234 & 0.2530 & 0.3963 \\ \hline
 0.76 & 0.2266 & 0.2653 & 0.3944 \\ \hline
 0.84 & 0.8382 & 0.1714 & 0.6506 \\ \hline \hline
 \end{tabular}
 \end{center}
 \vspace{-1mm}  
 \caption{\label{tab1} Some examples of the acceptable SFH models with the
 sensitivity model parameter $\log F_{\nu\cmn {\rm th}}^{\rm max}$, and
 three $p_{_{\rm KS}}$ for the $\log F_\nu$, $\log E$ and $\rm DM_E$
 criteria against the CHIME/FRB data for 390 type~I FRBs
 outside regions~(3). The boldfaced ones are also presented in
 the accompanying plots. See Sec.~\ref{sec3b} for details.}
 \end{table}



 \begin{center}
 \begin{figure}[tb]
 \centering
 \includegraphics[width=0.98\textwidth]{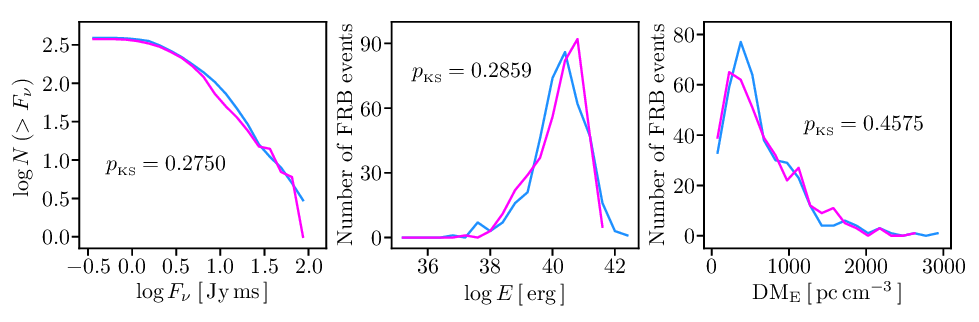}
 \vspace{-1mm}  
 \caption{\label{fig5} The simulated SFH model with $\log F_{\nu\cmn
 {\rm th}}^{\rm max}=0.77$ (magenta lines) against the CHIME/FRB data
 for 390 type~I FRBs outside regions~(3) (dodgerblue lines), with respect
 to the $\log F_\nu$, $\log E$ and $\rm DM_E$ criteria. Note that the
 simulations are all scaled to the CHIME/FRB data. The p-values
 $p_{_{\rm KS}}$ of the KS tests are given, respectively. See
 Sec.~\ref{sec3b} for details.}
 \end{figure}
 \end{center}


\vspace{-9mm} 

However, these $N_{\rm sim}$ mock FRBs intrinsically generated above are
 not the ones ``\,detected\,'' by the telescope, due to the telescope's
 sensitivity threshold and instrumental selection effects near
 the threshold. Therefore, the next step is to filter them by using the
 telescope's sensitivity model, which is difficult to characterize in fact.
 Here, we consider the simplified sensitivity model for CHIME following
 \cite{Zhang:2021kdu,Qiang:2021ljr}, in which the sensitivity threshold is
 about $0.3\;{\rm Jy\,ms}$ for CHIME, or equivalently $\log F_{\nu\cmn
 {\rm min}}=-0.5$, where the specific fluence is in units of Jy\,ms. Due to
 the direction-dependent sensitivity of the telescope, there is a ``\,gray
 zone\,'' in the $\log F_\nu$ distribution, within which CHIME has not
 reached full sensitivity to all sources~\cite{Zhang:2021kdu,Qiang:2021ljr}.
 The detection efficiency parameter in the ``\,gray zone\,'' is given by
 $\eta_{\rm det}={\cal R}^3$, where ${\cal R}=(\log F_{\nu\cmn {\rm th}}
 -\log F_{\nu\cmn {\rm th}}^{\rm min})/(\log F_{\nu\cmn {\rm th}}^{\rm max}
 -\log F_{\nu\cmn {\rm th}}^{\rm min})$, such that $\eta_{\rm det}\to 0$
 at $\log F_{\nu\cmn {\rm th}}^{\rm min}=-0.5$ and $\eta_{\rm det}\to 1$ at
 $\log F_{\nu\cmn {\rm th}}^{\rm max}$. Outside the ``\,gray zone\,'',
 $\eta_{\rm det}=1$. The filtered mock sample of FRBs will be confronted
 with the observational data, by using the KS tests with respect to the
 $\log F_\nu$, $\log E$ and $\rm DM_E$ distributions. In our case of SFH,
 the only free parameter is $\log F_{\nu\cmn {\rm th}}^{\rm max}$, which
 should be adjusted to match the observation. Notice that there are a few
 minor differences between the present work and~\cite{Qiang:2021ljr}. Here,
 we use slightly different $\rm DM_{host}$ and $f_{\rm IGM}$, while we have
 required the very conservative cut $\rm DM_{IGM}\geq DM_{obs}/10$ to
 exclude the actual bursts very close to us in the first CHIME/FRB data
 table, as in Sec.~\ref{sec2}. In practice, most of these $N_{\rm sim}$ mock
 FRBs cannot pass the filter of sensitivity threshold and instrumental
 selection effects. To ensure that there are still enough mock FRBs ($\sim
 {\cal O}(10^2)$, comparable with the number of observed FRBs) after the
 filter of sensitivity threshold and instrumental selection effects, we
 generate $N_{\rm sim}=4,000,000$ mock FRBs in the simulation for our case
 of SFH. We strongly refer to~\cite{Qiang:2021ljr} for the technical
 details.

Now, we test regions $(1)\sim (7)$ in the $\nu W-L_\nu$ phase plane one by
 one. For each region, 430 type~I FRBs are divided into two samples inside
 or outside this region. We compare their redshift distributions by using KS
 test, and also check whether one of these two sample tracks SFH by using
 the above method closely following~\cite{Qiang:2021ljr}. For 6 of these
 7 regions, although their redshift distributions are fairly different,
 both samples of type~I FRBs inside/outside the given region do not track
 SFH. The only survivor is region~(3), which is not so bad in some sense.
 It is defined by three physical conditions simultaneously
 \be{eq11}
 {\rm Region\ (3):}\quad\nu W\leq 10^{-3}\,{\rm GHz\;s}
 \quad\&\quad L_\nu\leq 10^{34}\,{\rm erg/s/Hz}\quad\&\quad
 T_B\geq 2\times 10^{35}\,{\rm K}\,.
 \ee
 As shown in left panel of Fig.~\ref{fig4}, there are 40 (390) type~I
 FRBs inside (outside) region~(3), and their redshift distributions are
 significantly different, with a p-value $p_{_{\rm KS}}=1.51\times 10^{-8}
 \ll 0.05$. We find that the 40 type~I FRBs inside region~(3) do not track
 SFH. However, the 390 type~I FRBs outside region~(3) can be consistent with
 SFH. In Table~\ref{tab1}, we show some examples of the acceptable SFH
 models with the sensitivity model parameter $\log F_{\nu\cmn {\rm th}}^{\rm
 max}$, and three $p_{_{\rm KS}}$ for the $\log F_\nu$, $\log E$ and
 $\rm DM_E$ criteria against the CHIME/FRB data for 390 type~I FRBs outside
 regions~(3). For some suitable $\log F_{\nu\cmn {\rm th}}^{\rm max}$, three
 p-values $p_{_{\rm KS}}>0.25$ simultaneously. We present an explicit
 example with $\log F_{\nu\cmn {\rm th}}^{\rm max}=0.77$ in Fig.~\ref{fig5},
 whose three p-values $p_{_{\rm KS}}>0.27$ simultaneously. So, SFH cannot
 be rejected at high confidence by the CHIME/FRB data for 390 type~I
 FRBs outside regions~(3), although three p-values are not fairly high.


\begin{table}[tb]
 \renewcommand{\arraystretch}{1.7}
 \begin{center}
 \vspace{-8mm}  
 \begin{tabular}{ccccc} \hline\hline
 \quad $\log F_{\nu\cmn {\rm th}}^{\rm max}$ \quad
 & \quad $p_{_{\rm KS}}$ for $\log F_\nu$ \quad
 & \quad $p_{_{\rm KS}}$ for $\log E$ \quad
 & \quad $p_{_{\rm KS}}$ for $\rm DM_E$ \quad \\ \hline
 {\bf 0.91} & {\bf 0.9621} & {\bf 0.5015} & {\bf 0.6550} \\ \hline
 0.96 & 0.7832 & 0.5050 & 0.5656 \\ \hline
 0.94 & 0.8742 & 0.4341 & 0.7663 \\ \hline
 0.87 & 0.7526 & 0.4238 & 0.7071 \\ \hline \hline
 \end{tabular}
 \end{center}
 \vspace{-1mm}  
 \caption{\label{tab2} The same as in Table~\ref{tab1}, but for 365
 type~I FRBs outside regions~(8).}
 \end{table}



 \begin{center}
 \begin{figure}[tb]
 \centering
 \includegraphics[width=0.98\textwidth]{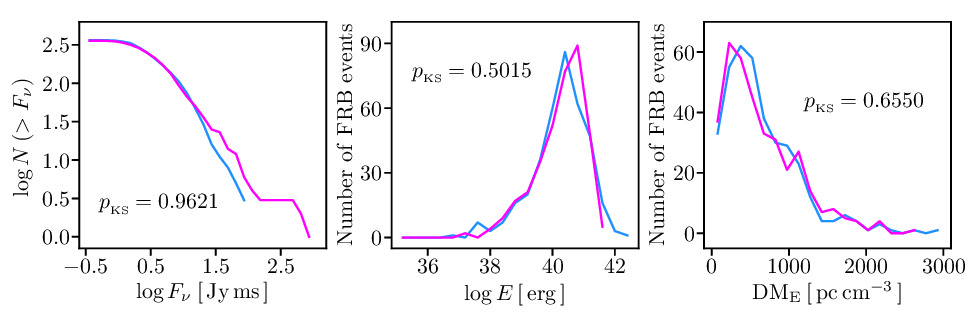}
 \vspace{-1mm}  
 \caption{\label{fig6} The same as in Fig.~\ref{fig5}, but for $\log F_{\nu
 \cmn {\rm th}}^{\rm max}=0.91$ and 365 type~I FRBs outside regions~(8).}
 \end{figure}
 \end{center}


\vspace{-10.7mm} 

Could we improve these results? The answer is a loud yes. Let us come back
 to Fig.~3. From its panels (a), (b) and (c), we easily find that the
 p-value for the dividing line with respect to $\nu W$ ($\sim 2\times
 10^{-5}$) is much larger than the ones with respect to $L_\nu$ ($\sim
 4\times 10^{-59}$) and $T_B$ ($\sim 10^{-15}$). Thus, it is reasonable
 to discard the dividing line $\nu W=10^{-3}\,{\rm Ghz\;s}$ from the
 $\nu W-L_\nu$ phase plane to improve the situation. In this case, the
 $\nu W-L_\nu$ phase plane is divided into 4 regions as shown in panel (d)
 of Fig.~3, namely region~(8) = regions $(1)+(3)$, region~(9) = regions
 $(4)+(5)$, region~(10) = regions $(2)+(7)$, and region~(6). Again, we
 test these 4 regions one by one. We find that the last three regions
 fail. However, region~(8) is very successful, which is defined by two
 physical conditions simultaneously
 \be{eq12}
 {\rm Region\ (8):}\quad L_\nu\leq 10^{34}\,{\rm erg/s/Hz}\quad\&\quad
 T_B\geq 2\times 10^{35}\,{\rm K}\,.
 \ee
 As shown in right panel of Fig.~\ref{fig4}, there are 65 (365) type~I
 FRBs inside (outside) region~(8), and their redshift distributions are
 significantly different, with a p-value $p_{_{\rm KS}}=5.13\times 10^{-12}
 \ll 0.05$ (note that this $p_{_{\rm KS}}$ is also much smaller than the
 one for region~(3), namely $1.51\times 10^{-8}$). We find that the 65
 type~I FRBs inside region~(8) do not track SFH. But the 365 type~I FRBs
 outside region~(8) do track SFH at high confidence. In Table~\ref{tab2}, we
 show some examples of the acceptable SFH models with the sensitivity model
 parameter $\log F_{\nu\cmn {\rm th}}^{\rm max}$, and three $p_{_{\rm KS}}$
 for the $\log F_\nu$, $\log E$ and $\rm DM_E$ criteria against the
 CHIME/FRB data for 365 type~I FRBs outside regions~(8). Clearly, for some
 suitable $\log F_{\nu\cmn {\rm th}}^{\rm max}$, three p-values $p_{_{\rm
 KS}}>0.5$ simultaneously. We present an explicit example with $\log F_{\nu
 \cmn {\rm th}}^{\rm max}=0.91$ in Fig.~\ref{fig6}. Thus, SFH can be fully
 consistent with the CHIME/FRB data for 365 type~I FRBs outside regions~(8).


 \begin{center}
 \begin{figure}[tb]
 \centering
 \vspace{-17mm}  
 \hspace{-5mm}  
 \includegraphics[width=0.85\textwidth]{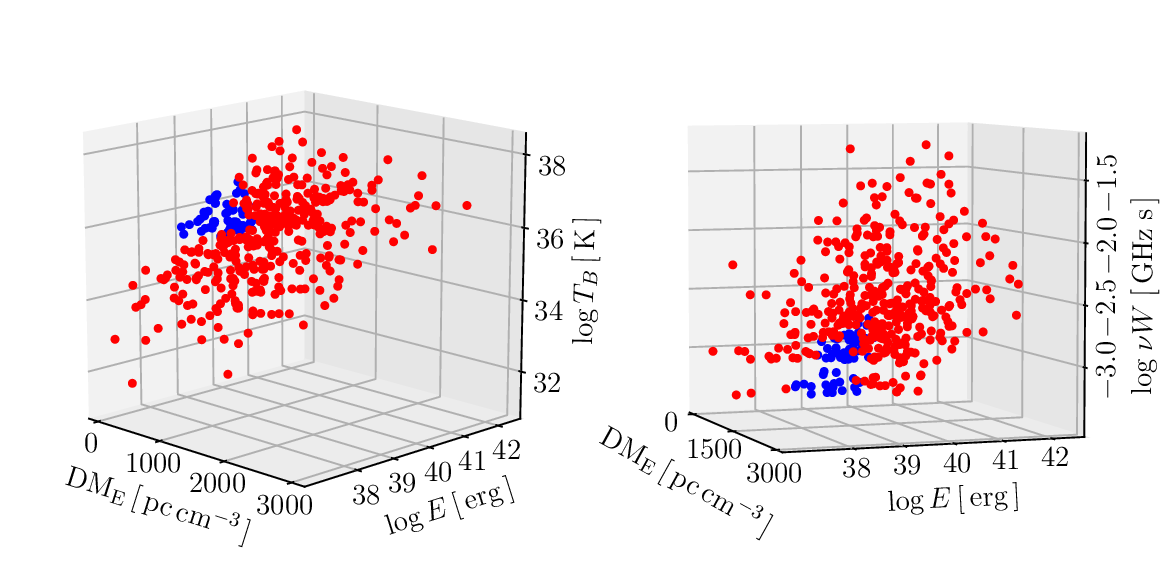}
 \caption{\label{fig7} Type~Ia FRBs (blue points) and type~Ib FRBs (red
 points) are located in distinct regions of 3-D spaces ${\rm DM_E}-\log
 E-\log T_B$ (left panel) or ${\rm DM_E}-\log E-\log\nu W$ (right panel).
 See Sec.~\ref{sec4} for details.}
 \end{figure}
 \end{center}



 \begin{center}
 \begin{figure}[tb]
 \centering
 \vspace{1mm}  
 \hspace{-5mm}  
 \includegraphics[width=0.98\textwidth]{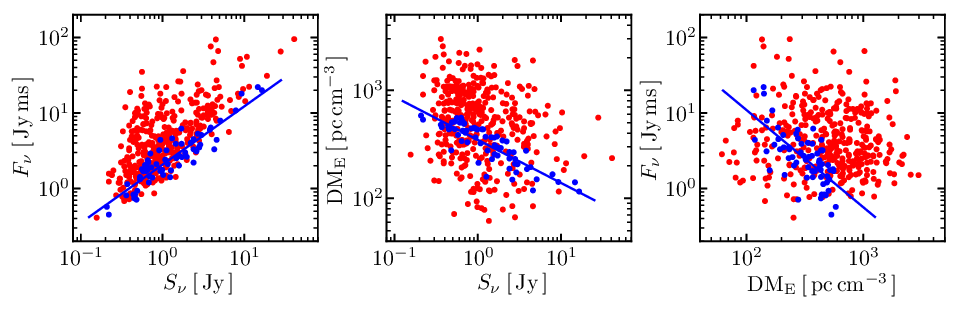}
 \caption{\label{fig8} The empirical correlations (blue lines) between
 fluence $F_\nu$, flux $S_\nu$ and $\rm DM_E$ for 65 type~Ia FRBs (blue
 points), but no such correlations for 365 type~Ib FRBs (red points).
 See Sec.~\ref{sec4} for details.}
 \end{figure}
 \end{center}


\vspace{-18.4mm} 

So far, we have identified a special region~(8) in the $\nu W-L_\nu$
 phase plane, which is defined by two physical conditions in
 Eq.~(\ref{eq12}) simultaneously. 430 type~I FRBs are divided into two
 distinct samples. The 65 type~I FRBs inside region~(8) do not track SFH,
 and hence they are probably associated with old stellar populations. We
 suggest calling them type~Ia FRBs. On the other hand, the 365 type~I FRBs
 outside region~(8) do track SFH, and hence they are probably associated
 with young stellar populations. We suggest calling them type~Ib FRBs. In
 this way, we have achieved a physical subclassification of type~I FRBs.
 From right panel of Fig.~\ref{fig4}, it is easy to see that type~Ib FRBs
 can appear at very high redshifts up to $z\sim 3.7$, but type~Ia FRBs can
 only be triggered at fairly low redshifts $z\lesssim 0.7$. A delay is
 required for type~Ia FRBs with respect to type~Ib FRBs (which track SFH).
 Quite different physical mechanisms are necessary for type Ia and Ib
 FRBs, respectively.


\section{Discriminating properties}\label{sec4}

After the subclassification of type~I FRBs, we would like to see their
 discriminating physical properties. It is worth noting that the only
 consideration in the previous section is whether one subclass of type~I
 FRBs tracks SFH while the other subclass does not, namely we completely
 have not taken their physical properties into account when we made this
 subclassification. We identified type~Ia and Ib FRBs only in the light
 of SFH. Logically, if they do come from different physical mechanisms,
 their physical properties might be also different in some aspects.


 \begin{center}
 \begin{figure}[tb]
 \centering
 \vspace{-16mm}  
 \hspace{-5mm}  
 \includegraphics[width=0.85\textwidth]{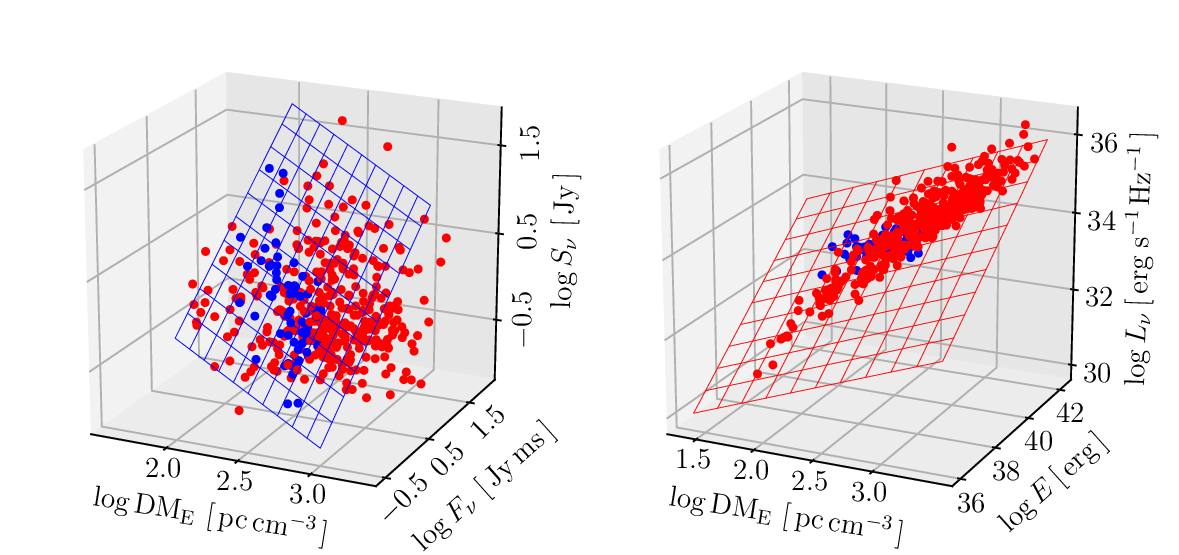}
 \caption{\label{fig9} Left panel: The 3-D empirical correlation (blue
 meshed plane) between fluence $F_\nu$, flux $S_\nu$ and $\rm DM_E$ given
 in Eq.~(\ref{eq16}) for 65 type~Ia FRBs (blue points), but no such
 correlation for 365 type~Ib FRBs (red points). Right panel: The 3-D
 empirical correlation (red meshed plane) between spectral
 luminosity $L_\nu$, isotropic energy $E$ and $\rm DM_E$ given
 in Eq.~(\ref{eq27}) for 365 type~Ib FRBs (red points). The ones for 65
 type~Ia FRBs (blue points) and 430 type~I FRBs (all points) are given in
 Eqs.~(\ref{eq26}) and (\ref{eq28}), respectively. See Sec.~\ref{sec4} for
 details.}
 \end{figure}
 \end{center}



 \begin{center}
 \begin{figure}[tb]
 \centering
 \vspace{1mm}  
 \hspace{-5mm}  
 \includegraphics[width=0.98\textwidth]{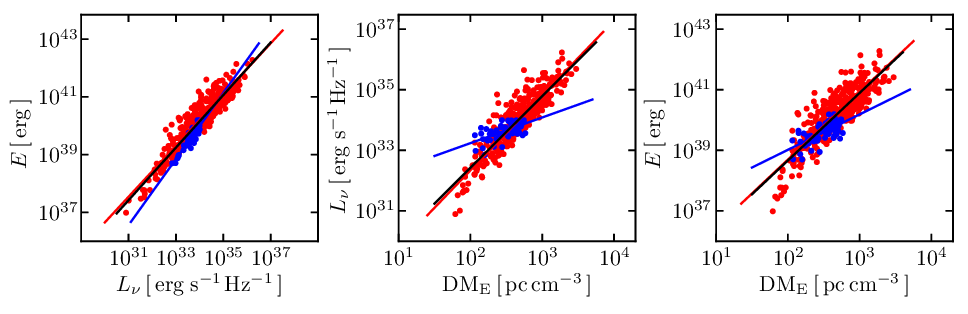}
 \caption{\label{fig10} The empirical correlations (blue/red/black lines)
 between spectral luminosity $L_\nu$, isotropic energy $E$ and $\rm DM_E$
 for 65 type~Ia FRBs (blue points), 365 type~Ib FRBs (red points), and 430
 type~I FRBs (all points), respectively. See Sec.~\ref{sec4} for details.}
 \end{figure}
 \end{center}


\vspace{-18.4mm} 

In Fig.~\ref{eq1} and panel~(d) of Fig.~\ref{fig3}, by definition, type~Ia
 and Ib FRBs are clearly separated in the $\nu W-L_\nu$ phase plane. Type~Ia
 FRBs have relatively high brightness temperatures $T_B$ and low spectral
 luminosities $L_\nu$, as required by Eq.~(\ref{eq12}). It is worth noting
 that we do not take transient duration $\nu W$ into account when the
 physical conditions in Eq.~(\ref{eq12}) are determined. However, an upper
 boundary in $\nu W$ naturally emerges for type~Ia FRBs, namely $\nu W
 \lesssim 2\times 10^{-3}\,{\rm GHz\;s}$, which is determined by the right
 vertex of region~(8) in Fig.~\ref{eq1} and panel (d) of Fig.~\ref{fig3}.
 So, type~Ia FRBs have also relatively short transient durations $\nu W$.
 Type~Ib FRBs are the rest outside region~(8). These two subclasses are
 clearly separated in this 2-D phase plane. Although they overlap each
 other in the ${\rm DM_E}-\log E$ plane, we find that type~Ia and Ib FRBs
 are located in distinct regions if the third dimension $\log T_B$ or
 $\log\nu W$ is added, as shown in Fig.~\ref{fig7}. Their separation in 3-D
 spaces also hints different physical mechanisms.

We try to find some empirical correlations for type~Ia and Ib FRBs through
 trial and error. At first, we present the empirical correlations between
 fluence $F_\nu$, flux $S_\nu$ and $\rm DM_E$. In Fig.~\ref{fig8}, we show
 the 2-D plots for these empirical correlations. Since our main purpose is
 to find the discriminating physical properties for type~Ia and Ib FRBs,
 it is enough to fit the data without error bars (and we will take errors
 into account in the future works). This can be done by using
 {\tt sklearn.linear$_{-}$model.LinearRegression} in
 Python~\cite{LinearRegression}. The score (coefficient of determination) is
 given by $R^2\equiv 1-\sum_k\left(y_k-\hat{y}_k\right)^2/\sum_k\left(y_k-
 \bar{y}\right)^2$, where $y_k$, $\hat{y}_k$ and $\bar{y}$ are
 the observed values, regressed values and mean of observed
 values~\cite{LinearRegression,Xiao:2021viy}, respectively. The higher $R$
 indicates the better fit, and $R=1$ at best. As shown in Fig.~\ref{fig8},
 we find tight 2-D empirical correlations for 65 type~Ia FRBs, namely
 \bea
 &&\log F_\nu=0.7709\log S_\nu+0.3150\,,\quad {\rm with}\quad
 R=0.9119\,,\label{eq13}\\[1mm]
 &&\log {\rm DM_E}=-0.3987\log S_\nu+2.5402\,,\quad {\rm with}\quad
 R=0.8626\,,\label{eq14}\\[1mm]
 &&\log F_\nu=-1.2873\log {\rm DM_E}+3.6139\,,\quad {\rm with}
 \quad R=0.7038\,.\label{eq15}
 \eea
 On the contrary, there are no such correlations for 365 type~Ib FRBs (since
 the corresponding $R^2<0$), as expected by eyes. Putting
 Eqs.~(\ref{eq13})\,--\,(\ref{eq15}) together, it is anticipated that there
 is a tight 3-D empirical correlation between fluence $F_\nu$, flux
 $S_\nu$ and $\rm DM_E$ for 65 type~Ia FRBs. Fitting to the data, we find
 \be{eq16}
 \log S_\nu=-0.9468\log {\rm DM_E}+0.7143\log F_\nu+2.1880\,,
 \quad {\rm with}\quad R=0.9634\,,
 \ee
 which is a 2-D plane in the 3-D plot as shown in left panel of
 Fig.~\ref{fig9}. Noting that its $R$ is much higher than the ones of
 2-D empirical correlations given in Eqs.~(\ref{eq13})\,--\,(\ref{eq15}),
 we recommend preferably using the tight 3-D empirical correlation given
 in Eq.~(\ref{eq16}). No such 3-D correlation for 365 type~Ib FRBs.

On the other hand, we also find some empirical correlations between spectral
 luminosity $L_\nu$, isotropic energy $E$ and $\rm DM_E$. As shown by the
 blue lines in Fig.~\ref{fig10}, we find the 2-D empirical correlations
 for 65 type~Ia FRBs, namely
 \bea
 &&\log E=1.1446\log L_\nu+1.0720\,,\quad {\rm with}\quad
 R=0.9081\,,\label{eq17}\\[1mm]
 &&\log L_\nu=0.8470\log {\rm DM_E}+31.5430\,,\quad {\rm with}\quad
 R=0.6112\,,\label{eq18}\\[1mm]
 &&\log E=1.1698\log {\rm DM_E}+36.6770\,,\quad {\rm with}
 \quad R=0.6697\,.\label{eq19}
 \eea
 There are similar 2-D empirical correlations for 365 type~Ib FRBs as shown
 by the red lines in Fig.~\ref{fig10}, but with quite different slopes and
 intercepts, namely
 \bea
 &&\log E=0.8862\log L_\nu+10.0664\,,\quad {\rm with}\quad
 R=0.9285\,,\label{eq20}\\[1mm]
 &&\log L_\nu=2.4707\log {\rm DM_E}+27.3976\,,\quad {\rm with}\quad
 R=0.9065\,,\label{eq21}\\[1mm]
 &&\log E=2.2345\log {\rm DM_E}+34.2238\,,\quad {\rm with}
 \quad R=0.8590\,.\label{eq22}
 \eea
 Obviously, these fits are much better than the ones for 65 type~Ia FRBs,
 since they have much higher $R$. If we instead consider all 430 type~I
 FRBs as a whole, these 2-D empirical correlations become
 \bea
 &&\log E=0.9106\log L_\nu+9.1906\,,\quad {\rm with}\quad
 R=0.9249\,,\label{eq23}\\[1mm]
 &&\log L_\nu=2.3656\log {\rm DM_E}+27.6917\,,\quad {\rm with}\quad
 R=0.8952\,,\label{eq24}\\[1mm]
 &&\log E=2.2309\log {\rm DM_E}+34.2027\,,\quad {\rm with}
 \quad R=0.8575\,,\label{eq25}
 \eea
 as shown by the black lines in Fig.~\ref{fig10}. Obviously, they are very
 close to the ones for 365 type~Ib FRBs. It is not surprising since 365
 type~Ib FRBs dominate the whole type~I sample. In the light of the 2-D
 empirical correlations in Eqs.~(\ref{eq17})\,--\,(\ref{eq25}), it is
 anticipated that there is a tight 3-D empirical correlation between
 spectral luminosity $L_\nu$, isotropic energy $E$ and $\rm DM_E$. Fitting
 to the data, we find
 \bea
 &{\rm Type~Ia:}\quad &\log L_\nu=0.0079\log {\rm DM_E}+0.7174\log E+5.2316
 \,,\quad {\rm with}\quad R=0.9081\,,\label{eq26}\\[1mm]
 &{\rm Type~Ib:}\quad &\log L_\nu=1.1330\log {\rm DM_E}+0.5986\log E+6.9098
 \,,\quad {\rm with}\quad R=0.9525\,,\label{eq27}\\[1mm]
 &{\rm Type~I:}\quad &\log L_\nu=1.0196\log {\rm DM_E}+0.6033\log E+7.0561
 \,,\quad {\rm with}\quad R=0.9460\,,\label{eq28}
 \eea
 for 65 type~Ia, 365 type~Ib and all 430 type~I FRBs, respectively. They
 are much better than the ones of 2-D empirical correlations in
 Eqs.~(\ref{eq17})\,--\,(\ref{eq25}), since they have much higher $R$.
 We recommend preferably using the tight 3-D empirical correlations given
 in Eqs.~(\ref{eq26})\,--\,(\ref{eq28}). In right panel of Fig.~\ref{fig9},
 we present the 2-D plane corresponding to Eq.~(\ref{eq27}) in the 3-D
 plot. Although they have similar empirical correlations, type~Ia and Ib
 FRBs are still distinguishable, due to their quite different slopes
 and intercepts.

So far, we show that type~Ia and Ib FRBs have some discriminating physical
 properties. They can be clearly separated in the $\nu W-L_\nu$ phase plane
 and some 3-D spaces. We find some tight empirical correlations for them,
 which can also be used to distinguish between type~Ia and Ib FRBs. These
 results hint that they do come from different physical mechanisms.


\begin{table}[tb]
 \renewcommand{\arraystretch}{1.7}
 \begin{center}
 \vspace{-7mm}  
 \begin{tabular}{ccccc} \hline\hline
 \quad $\log F_{\nu\cmn {\rm th}}^{\rm max}$ \quad
 & \quad $p_{_{\rm KS}}$ for $\log F_\nu$ \quad
 & \quad $p_{_{\rm KS}}$ for $\log E$ \quad
 & \quad $p_{_{\rm KS}}$ for $\rm DM_E$ \quad \\ \hline
 {\bf 0.95} & {\bf 0.9180} & {\bf 0.3951} & {\bf 0.7758} \\ \hline
 0.96 & 0.8801 & 0.3928 & 0.7769 \\ \hline
 0.97 & 0.8504 & 0.3911 & 0.7120 \\ \hline
 0.99 & 0.7631 & 0.3868 & 0.6610 \\ \hline
 1.0 & 0.7036 & 0.3841 & 0.6185 \\ \hline \hline
 \end{tabular}
 \end{center}
 \vspace{-1mm}  
 \caption{\label{tab3} The same as in Table~\ref{tab1}, but for 365
 type~Ib FRBs + 17 repeaters.}
 \end{table}



 \begin{center}
 \begin{figure}[tb]
 \centering
 \includegraphics[width=0.98\textwidth]{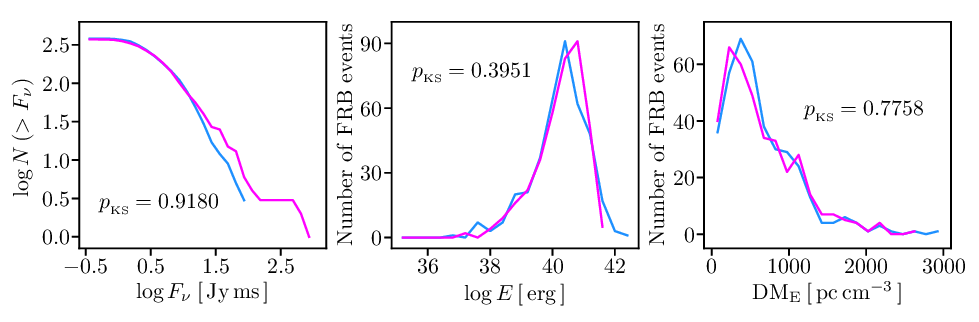}
 \vspace{-1mm}  
 \caption{\label{fig11} The same as in Fig.~\ref{fig5}, but for $\log F_{\nu
 \cmn {\rm th}}^{\rm max}=0.95$ and 365 type~Ib FRBs + 17 repeaters.}
 \end{figure}
 \end{center}


\vspace{-11mm} 


\section{Repeating FRBs}\label{sec5}

Let us turn to repeating (type~II) FRBs. It is reasonable to also consider a
 similar subclassification of type~II FRBs. As mentioned in Sec.~\ref{sec2},
 there are only 17 repeaters after the robust cut, which are too few to
 form a good enough sample in statistics. But this does not prevent us
 from some exploratory studies.

The first question is whether there are subclasses of repeating
 (type~II) FRBs? The answer is a clear yes. In the recent observations, some
 extragalactic repeaters were located in the star-forming environments, as
 mentioned in Sec.~\ref{sec1}. The Galactic repeater FRB 200428 has been
 firmly associated with the young magnetar SGR 1935+2154. On the other hand,
 the well-known repeating FRB 20200120E in a globular cluster is clearly
 associated with old stellar populations. These observations strongly
 suggest a similar subclassification of repeating (type~II) FRBs: type~IIa
 FRBs (e.g.~FRB 20200120E) are associated with old stellar populations and
 hence do not track SFH, while type~IIb FRBs (e.g.~FRB 180916.J0158+65, FRB
 121102, FRB 20190520B, FRB 20181030A, FRB 200428) are associated with young
 stellar populations and hence track SFH. Different physical mechanisms are
 required by type~IIa and IIb FRBs.

The second question is whether there are two subclasses in the 17 repeaters
 of the first CHIME/FRB catalog? The answer might be not. It is reasonable
 to speculate that they are all type~IIb FRBs, because (a) the known
 type~IIa repeater associated with old stellar populations (i.e.~FRB
 20200120E) is not in the first CHIME/FRB catalog. (b) at least 16 of these
 17 repeaters are clearly outside region~(8) in the $\nu W-L_\nu$ phase
 plane as shown in Fig.~\ref{fig1}, while the only one just at the right
 vertex of region~(8) can also be excluded due to the uncertainty of the
 exact boundaries of region~(8). (c) at least two of the known type~IIb
 repeaters associated with young stellar populations (i.e.~FRB 20121102A and
 FRB 20180916B (FRB 180916.J0158+65)) are in the first CHIME/FRB catalog,
 namely they are 2 of the 17 repeaters under discussion. (d) if these 17
 repeaters are all type~IIb FRBs, we could check this speculation by
 combining them with 365 type~Ib FRBs and see whether these 382 FRBs track
 SFH (n.b.~17 repeaters are too few to do this alone since they cannot form
 a good enough sample in statistics). This can be tested by using the method
 mentioned in Sec.~\ref{sec3b}, and we present the results in
 Table~\ref{tab3} and Fig.~\ref{fig11}. Clearly, these 382 FRBs can be
 fully consistent with SFH since three p-values $p_{_{\rm KS}}>0.39$
 simultaneously for some suitable $\log F_{\nu\cmn {\rm th}}^{\rm max}$. So,
 the possibility that these 17 repeaters are all type~IIb FRBs (associated
 with young stellar populations and hence track SFH) cannot be excluded
 by now.

The third question is that could we speculate the physical criteria for the
 subclassification of type~II FRBs? Let us try. It is natural to also
 subclassify type~II FRBs in the $\nu W-L_\nu$ phase plane, and the
 conservative physical criteria might be similar to the ones of type~I FRBs,
 namely they might also be the dividing lines of $L_\nu$ and $T_B$ (and
 $\nu W$). But it is reasonable that the dividing lines of $L_\nu$ and $T_B$
 (and $\nu W$) might be quite different from the ones of type~I FRBs (namely
 $L_{\nu,\,\rm crit}=10^{34}\,{\rm erg/s/Hz}$ and $T_{B,\,\rm crit}=2\times
 10^{35}\,{\rm K}$). For instance, a much lower $T_{B,\,\rm crit}=10^{33}\,
 {\rm K}$ was proposed for FRB 20121102A in~\cite{Xiao:2021viy}, although
 their main purpose is somewhat different from ours. Here, we would like to
 make a bold but not too bold speculation. We strongly refer to Fig.~3
 of~\cite{Nimmo:2021yob} or Fig.~7 of~\cite{Petroff:2021wug}, where the
 known repeater associated with old stellar populations (type~IIa), i.e.~FRB
 20200120E, is plotted in the $\nu W-L_\nu$ phase plane, together with the
 known repeaters associated with young stellar populations (type~IIb),
 i.e.~FRB 20180916B (FRB 180916.J0158+65), FRB 20121102A, FRB 20190711A,
 and the Galactic FRB 200428 (SGR 1935+2154). Thus, Fig.~3
 of~\cite{Nimmo:2021yob} or Fig.~7 of~\cite{Petroff:2021wug} are ideal test
 ground for the subclassification of type~II FRBs. Similar to region~(8)
 defined by Eq.~(\ref{eq12}) for type~I FRBs, we speculate that the possible
 physical criteria for the subclassification of type~II FRBs might be
 given by
 \bea
 &{\rm Type~IIa:}\quad & L_\nu\lesssim 10^{29}\,{\rm erg/s/Hz}\quad
 \&\quad T_B\gtrsim 10^{30}\,{\rm K}\,,\label{eq29}\\[1mm]
 &{\rm Type~IIb:}\quad & {\rm otherwise}\,.\label{eq30}
 \eea
 Note that they are roughly estimated by eyes from Fig.~3
 of~\cite{Nimmo:2021yob} or Fig.~7 of~\cite{Petroff:2021wug}, and hence
 they are not the exact ones. In this way, FRB 20200120E, the known type~IIa
 repeater associated with old stellar populations, could be roughly
 separated from the known type~IIb repeaters associated with young stellar
 populations mentioned above. If these physical criteria for
 the subclassification of type~II FRBs given by Eqs.~(\ref{eq29}) and
 (\ref{eq30}) are roughly correct, it is easy to see from Fig.~\ref{fig1}
 that the 17 CHIME repeaters are all type~IIb FRBs, coincident with
 our discussions about the second question.

Although the number of repeaters under consideration is too few to go
 further, we have tried our best to subclassify the repeaters into type~IIa
 and IIb FRBs, with good enough reasonings based on the observational facts.
 We stress that they are highly speculative. Since the data of repeaters
 will be rapidly accumulated in the future, we hope this subclassification
 of type~II FRBs could be refined.


\section{Concluding remarks}\label{sec6}

Although FRBs have been an active field in astronomy and cosmology, their
 origin is still unknown to date. One of the interesting topics is the
 classification of FRBs, which is closely related to the origin of FRBs.
 Different physical mechanisms are required by different classes of FRBs.
 In the literature, they usually could be classified into non-repeating
 and repeating FRBs. Well motivated by the observations, here we are
 interested in the possible subclassification of FRBs. By using the first
 CHIME/FRB catalog, we propose to subclassify non-repeating (type~I) FRBs
 into type Ia and Ib FRBs. The distribution of type~Ia FRBs is delayed with
 respect to SFH, and hence they are probably associated with old stellar
 populations, while the distribution of type~Ib FRBs tracks SFH, and hence
 they are probably associated with young stellar populations. Accordingly,
 the physical criteria for this subclassification of type~I FRBs have
 been clearly determined. We find that there are some tight empirical
 correlations for type~Ia FRBs but not for type~Ib FRBs, and vice versa.
 These make them different in physical properties. Similarly, we suggest
 that repeating (type~II) FRBs could also be subclassified into type~IIa
 and IIb FRBs. This subclassification of FRBs might help us to reveal
 quite different physical mechanisms behind them, and improve
 their applications in astronomy and cosmology.

In history, the subclassifications have made many important progresses in
 various fields. For example, the subclassification of supernovae is well
 known, as mentioned in Sec.~\ref{sec1}. The famous subclass of type~Ia
 supernovae was identified in this way. Only they can be used as standard
 candles, which led to the great discovery of cosmic acceleration (and
 Nobel prize in physics 2011). This highlights the importance of the
 subclassifications. It might happen again in the field of FRBs. For
 instance, FRBs can be used to study cosmology, but the constraints on
 the cosmological parameters are usually loose (see e.g.
 \cite{Deng:2013aga,Yang:2016zbm,Gao:2014iva,Zhou:2014yta,Qiang:2019zrs,
 Qiang:2020vta,Qiang:2021bwb}). If one of the subclasses of FRBs could be
 used as standard candles, rulers, or sirens (say, by the help of some
 unknown empirical correlations for this subclass of FRBs), we might
 remarkably improve the cosmological constraints in the future. Let
 us keep an open mind to this possibility.


\begin{table}[tb]
 \renewcommand{\arraystretch}{1.0}
 \begin{center}
 \vspace{-6mm}  
 \hspace{-5mm}  
 \begin{tabular}{c|c|c} \hline\hline
 {\bf FRBs} & \begin{tabular}{c} $\vspace{-3.7mm}$\\ {\bf Class (a)\;:} \\ associated with old stellar populations\\[1.298mm] \end{tabular}
 & \begin{tabular}{c} {\bf Class (b)\;:} \\ \ associated with young stellar populations \end{tabular} \\ \hline
 \begin{tabular}{c} {\bf Type I\;:} \\ Non-repeating\ \ \\ \end{tabular}
 & \begin{tabular}{c} $\vspace{-3.7mm}$\\ {\bf Type Ia\;:} \\ Non-repeating FRBs \\ associated with old stellar populations \\ \ and hence delayed with respect to SFH\;\ \\[1.298mm] \end{tabular}
 & \begin{tabular}{c} {\bf Type Ib\;:} \\ Non-repeating FRBs \\ \ associated with young stellar populations\ \\ and hence track SFH \end{tabular} \\ \hline
 \begin{tabular}{c} {\bf Type II\;:} \\ Repeating\ \ \end{tabular}
 & \begin{tabular}{c} $\vspace{-3.7mm}$\\ {\bf Type IIa\;:} \\ Repeating FRBs \\ associated with old stellar populations \\ \ and hence delayed with respect to SFH\;\ \\[1.298mm] \end{tabular}
 & \begin{tabular}{c} {\bf Type IIb\;:} \\ Repeating FRBs \\ \ associated with young stellar populations \\ and hence track SFH \end{tabular} \\ \hline \hline
 \end{tabular}
 \end{center}
 \vspace{-1mm}  
 \caption{\label{tab4} A brief summary of the universal
 subclassification scheme of FRBs.}
 \end{table}


In this work, we have identified 65 type~Ia FRBs in the first CHIME/FRB
 catalog. They have relatively high brightness temperatures $T_B$ and low
 spectral luminosities $L_\nu$, as required by Eq.~(\ref{eq12}). As shown
 in right panel of Fig.~\ref{fig4}, type~Ia FRBs can only be triggered at
 fairly low redshifts $z\lesssim 0.7$. A delay is required for type~Ia
 FRBs with respect to type~Ib FRBs (which track SFH). They are probably
 associated with old stellar populations. On the other hand, as mentioned
 at the beginning of Sec.~\ref{sec4}, an upper boundary in $\nu W$ naturally
 emerges for type~Ia FRBs, namely $\nu W\lesssim 2\times 10^{-3}\,{\rm
 GHz\;s}$. So, type~Ia FRBs have also relatively short transient durations
 $\nu W$. These properties might help us to reveal the physical mechanism
 for type~Ia FRBs. For instance, the compact binary merger model in a rapid
 process might be one of the candidates. Gravitational waves (GWs) are
 usually expected in such a merger. Thus, we might witness type~Ia FRBs as
 electromagnetic counterparts of GW events in the future. This might be a
 good chance to study gravity, cosmology and IGM.

As found in~\cite{Zhang:2021kdu,Qiang:2021ljr}, all FRBs in the first
 CHIME/FRB catalog as a whole do not track SFH. In this work, we have
 identified the main cause, namely 65 type~Ia FRBs. If they are removed,
 the rest (mainly type~Ib FRBs) do track SFH. In the future, numerous
 type~I FRBs could be well located in their host galaxies, and hence they
 could be easily subclassified: the ones associated with old/young stellar
 populations are type~Ia/Ib FRBs, respectively. So, one might only use
 type~Ib FRBs to study cosmology and IGM. Thus, it is justified that
 today one can generate the mock sample of type~Ib FRBs by simply assuming
 a redshift distribution tracking SFH, and use these mock type~Ib FRBs to
 study cosmology and IGM. In this way, one might avoid to consider the
 complicated redshift distribution models, and hence the simulations
 could be significantly simplified.

Currently, many of well located FRBs are repeaters~\cite{Heintz:2020},
 thanks to their repeating behaviors. So, it is reasonable to expect that
 there will be many observational data for repeating (type~II) FRBs well
 located in their host galaxies in the near future, and hence they could
 be easily subclassified: the ones associated with old/young stellar
 populations are type~IIa/IIb FRBs, respectively. At that time, the
 physical criteria (in terms of e.g. $L_\nu$, $T_B$, $\nu W$) for the
 subclassification of type~II FRBs could be clearly determined. In turn,
 they could be used to subclassify the repeaters without well located host
 galaxies. This virtuous cycle will benefit the studies on repeating
 (type~II) FRBs.

In Table~\ref{tab4}, we present a brief summary of the
 universal subclassification scheme of FRBs. As in the literature, type
 I/II FRBs are non-repeating/repeating, respectively. Class (a)/(b) FRBs
 are associated with old/young stellar populations, respectively. Their
 combinations result in four subclasses: Ia, Ib, IIa, IIb, as shown in
 Table~\ref{tab4}. The physical criteria for this subclassification given
 in Eqs.~(\ref{eq12}) and (\ref{eq29}) are inferred by using the first
 CHIME/FRB catalog and Fig.~3 of~\cite{Nimmo:2021yob} or Fig.~7
 of~\cite{Petroff:2021wug}, respectively. We stress that the physical
 criteria in Eqs.~(\ref{eq12}) and (\ref{eq29}) might be changed for
 the larger and better FRB datasets in the future, but the universal
 subclassification scheme given in Table~\ref{tab4} will always hold.
 Different physical mechanisms for FRBs are required by these subclasses.
 Note that the key improvement of the universal subclassification scheme
 given in Table~\ref{tab4} is that it works even for a single FRB. A sample
 of FRBs is needed to see whether their distribution tracks SFH, while one
 cannot determine whether a single FRB tracks SFH or not. But even for a
 single FRB, its host galaxy and local environment can be precisely
 determined. If this FRB has been localized in a star-forming environment,
 it is a class~(b) FRB. Otherwise, it is a class~(a) FRB. Combining with
 whether it repeats, we can then determine its subclass to be one of Ia,
 Ib, IIa, IIb. In this way, no other criteria are needed (on the other
 hand, for a single FRB without identified host galaxy, we could instead
 subclassify it by using the physical criteria given in Eq.~(\ref{eq12})
 or Eq.~(\ref{eq29})).

It is of interest to speculate the possible progenitor theories for these
 four subclasses of FRBs. We refer to~\cite{Platts:2018hiy} for the
 up-to-date FRB theory catalogue. In general, since class~(a) FRBs are
 associated with old stellar populations and hence delayed with respect to
 SFH, their progenitors might be formed via the compact binary merger which
 needs to undergo a long inspiral phase before the final coalescence, or the
 accretion-induced collapse of a white dwarf (WD) which also needs a long
 time to accrete before the final collapse. On the other hand, since
 class~(b) FRBs are associated with young stellar populations and hence
 track SFH, their progenitors might be formed directly via the collapse of
 a massive star. Therefore, we speculate that type~Ia FRB comes from the
 merger of neutron star (NS) $-$ black hole (BH) binary or NS-NS binary (see
 Sec.~4.1 of \cite{Platts:2018hiy}), and the final remnant of this merger is
 a black hole so that the resulted FRB is one-off. The progenitor of
 type~IIa FRB might be the magnetar formed via the merger of WD-WD binary
 or the accretion-induced collapse of WD (see e.g.~\cite{Lu:2021enm,
 Kirsten:2021llv}), while the magnetars similar to the well-known SGR
 1935+2154 are the leading progenitors for the repeating FRBs. Similarly,
 the progenitor of type~IIb FRB might be the young magnetar formed directly
 via the collapse of a massive star. The progenitor of type~Ib FRB might be
 the supramassive NS formed via the collapse of a massive star, and it
 quickly collapses into a black hole or a quark star to produce
 a one-off FRB (see Sec.~4.2 of \cite{Platts:2018hiy}). Of course, the above
 speculations are proposed just for examples. The other novel progenitor
 theories for these four subclasses of FRBs are all desirable.

We would like to briefly discuss the possible systematic errors. In
 generating the mock FRBs, the main systematic errors come from $\rm
 DM_{IGM}$. In principle, $\rm DM_{IGM}$ should deviate from the mean given
 in Eq.~(\ref{eq3}) if the plasma density fluctuations are taken into
 account~\cite{McQuinn:2013tmc,Ioka:2003fr,Jaroszynski:2018vgh}. In the
 literature, the typical error in $\rm DM_{IGM}$ is $\sigma_{\rm IGM}\sim
 100\,\dmunit$~\cite{McQuinn:2013tmc,Ioka:2003fr,
 Jaroszynski:2018vgh,Qiang:2021ljr,Deng:2013aga,Yang:2016zbm,Gao:2014iva,
 Zhou:2014yta,Qiang:2019zrs,Qiang:2020vta,Qiang:2021bwb}. However, we
 stress that it is not a serious problem in the present work, because we
 are simulating a very large sample of mock FRBs ($N_{\rm sim}=4,000,000$)
 and hence $\rm DM_{IGM}$ should heavily concentrate on the mean given in
 Eq.~(\ref{eq3})~\cite{Zhang:2021kdu}. Similarly, if we take the error of
 $\rm DM_{host}$ into account, it is also not a serious problem when we
 generate $\rm DM_E$ for a very large sample of mock FRBs ($N_{\rm sim}=
 4,000,000$)~\cite{Zhang:2021kdu}. On the other hand, our main goal is to
 subclassify FRBs, while the physical criteria given in Eq.~(\ref{eq12})
 or Eq.~(\ref{eq29}) are very rough in fact. Some uncertainties in the
 boundaries of e.g.~region~(8) in the $\nu W-L_\nu$ phase plane (see
 Fig.~\ref{fig1}) only affect a few observed FRBs, and hence cannot
 significantly change the subclassification of FRBs. Of course, they should
 be carefully considered in the future works.


\section*{ACKNOWLEDGEMENTS}

We thank the anonymous referee for quite useful comments and suggestions,
 which helped us to improve this work. We are grateful to Da-Chun~Qiang,
 Hua-Kai~Deng, Shu-Ling~Li, Shupeng~Song, Jing-Yi~Jia and Zi-Yu~Hou for
 kind help and useful discussions. This work was supported in part by NSFC
 under Grants No.~11975046 and No.~11575022.

\renewcommand{\baselinestretch}{1.12}



\begin{thebibliography}{99}

\bibitem{NAFRBs}
https:$/\!/$www.nature.com/collections/rswtktxcln

\bibitem{Lorimer:2018rwi}
  D.~R.~Lorimer,
  Nat.\ Astron.\  {\bf 2}, 860 (2018)
  [arXiv:1811.00195].

\bibitem{Keane:2018jqo}
  E.~F.~Keane,
  Nat.\ Astron.\  {\bf 2}, 865 (2018)
  [arXiv:1811.00899].

\bibitem{Caleb:2018ygr}
  M.~Caleb, L.~G.~Spitler and B.~W.~Stappers,
  Nat.\ Astron.\  {\bf 2}, 839 (2018)
  [arXiv:1811.00360].

\bibitem{Pen:2018ilo}
  U.~L.~Pen,
  Nat.\ Astron.\  {\bf 2}, 842 (2018)
  [arXiv:1811.00605].

\bibitem{Petroff:2019tty}
  E.~Petroff, J.~W.~T.~Hessels and D.~R.~Lorimer,
  Astron.\ Astrophys.\ Rev.\  {\bf 27}, 4 (2019)
  [arXiv:1904.07947].

\bibitem{Petroff:2021wug}
  E.~Petroff, J.~W.~T.~Hessels and D.~R.~Lorimer,
  Astron. Astrophys. Rev. \textbf{30}, 2 (2022)
  [arXiv:2107.10113].

\bibitem{Zhang:2020qgp}
  B.~Zhang,
  Nature {\bf 587}, 45 (2020)
  [arXiv:2011.03500].

\bibitem{Lyubarsky:2021bai}
  Y.~Lyubarsky,
  Universe \textbf{7}, no.3, 56 (2021)
  [arXiv:2103.00470].

\bibitem{Xiao:2021omr}
  D.~Xiao, F.~Y.~Wang and Z.~G.~Dai,
  Sci.\ China Phys.\ Mech.\ Astron.\  {\bf 64}, 249501 (2021)
  [arXiv:2101.04907].

\bibitem{Caleb:2021xqe}
  M.~Caleb and E.~Keane,
  Universe \textbf{7}, no.11, 453 (2021).

\bibitem{Nicastro:2021cxs}
  L.~Nicastro {\it et al.},
  Universe \textbf{7}, no.3, 76 (2021)
  [arXiv:2103.07786].

\bibitem{Cordes:2019cmq}
  J.~M.~Cordes and S.~Chatterjee,
  Ann.\ Rev.\ Astron.\ Astrophys.\  {\bf 57}, 417 (2019)
  [arXiv:1906.05878].

\bibitem{Pilia:2022mul}
  M.~Pilia,
  Universe \textbf{8}, 9 (2022)
  [arXiv:2203.04890].

\bibitem{Bhandari:2021thi}
  S.~Bhandari and C.~Flynn,
  Universe \textbf{7}, no.4, 85 (2021).

\bibitem{Platts:2018hiy}
  E.~Platts {\it et al.},
  Phys. Rept. \textbf{821}, 1 (2019)
  [arXiv:1810.05836].\\
  The up-to-date FRB
  theory catalogue is available at https:$/\!/$frbtheorycat.org

\bibitem{Petroff:2016tcr}
  E.~Petroff {\it et al.},
  Publ. Astron. Soc. Austral. \textbf{33}, e045 (2016)
  [arXiv:1601.03547].\\
  The up-to-date FRB Catalogue is available at https:$/\!/$www.frbcat.org
  and https:$/\!/$www.wis-tns.org

\bibitem{Heintz:2020}
  K.~E.~Heintz {\it et al.},
  Astrophys.\ J.\  {\bf 903}, 152 (2020)
  [arXiv:2009.10747].\\
  The up-to-date compilation of all known FRB host galaxies
  is available at https:$/\!/$frbhosts.org

\bibitem{Palaniswamy:2017aze}
  D.~Palaniswamy, Y.~Li and B.~Zhang,
  Astrophys. J. Lett. \textbf{854}, no.1, L12 (2018)
  [arXiv:1703.09232].

\bibitem{Ai:2020wnm}
  S.~Ai, H.~Gao and B.~Zhang,
  Astrophys. J. Lett. \textbf{906}, no.1, L5 (2021)
  [arXiv:2007.02400].

\bibitem{Caleb:2019szc}
  M.~Caleb {\it et al.},
  Mon. Not. Roy. Astron. Soc. \textbf{484}, 5500 (2019)
  [arXiv:1902.00272].

\bibitem{Connor:2020phs}
  L.~Connor {\it et al.},
  Mon. Not. Roy. Astron. Soc. \textbf{497}, 3076 (2020)
  [arXiv:2003.11930].

\bibitem{Bagchi:2017tzi}
  M.~Bagchi,
  Astrophys. J. Lett. \textbf{838}, no.2, L16 (2017)
  [arXiv:1702.08876].

\bibitem{Yamasaki:2017hdr}
  S.~Yamasaki, T.~Totani and K.~Kiuchi,
  Publ. Astron. Soc. Jap. \textbf{70}, no.3, 39 (2018)
  [arXiv:1710.02302].

\bibitem{Katz:2022cyt}
  J.~I.~Katz,
  arXiv:2203.03675 [astro-ph.HE].

\bibitem{Katz:2022sqt}
  J.~I.~Katz,
  Mon. Not. Roy. Astron. Soc. \textbf{513}, no.2, 1925 (2022)
  [arXiv:2201.02910].

\bibitem{Hashimoto:2020acj}
  T.~Hashimoto {\it et al.},
  Mon. Not. Roy. Astron. Soc. \textbf{498}, no.3, 3927 (2020)
  [arXiv:2008.09621].

\bibitem{Zhong:2022uvu}
  S.~Q.~Zhong {\it et al.},
  Astrophys. J. \textbf{926}, no.2, 206 (2022)
  [arXiv:2202.04422].

\bibitem{Pleunis:2021qow}
  Z.~Pleunis {\it et al.},
  Astrophys. J. \textbf{923}, no.1, 1 (2021)
  [arXiv:2106.04356].

\bibitem{Li:2021yds}
  X.~J.~Li, X.~F.~Dong, Z.~B.~Zhang and D.~Li,
  Astrophys. J. \textbf{923}, no.2, 230 (2021)
  [arXiv:2110.07227].

\bibitem{Xiao:2021viy}
  D.~Xiao and Z.~G.~Dai,
  Astron. Astrophys. \textbf{657}, L7 (2022)
  [arXiv:2112.12301].

\bibitem{Chaikova:2022vnh}
  A.~Chaikova, D.~Kostunin and S.~B.~Popov,
  arXiv:2202.10076 [astro-ph.HE].

\bibitem{SN}
  https:$/\!/$en.wikipedia.org/wiki/Supernova

\bibitem{Zhang:2006mb}
  B.~Zhang {\it et al.},
  Astrophys. J. Lett. \textbf{655}, L25 (2007)
  [astro-ph/0612238].

\bibitem{Kumar:2014upa}
  P.~Kumar and B.~Zhang,
  Phys. Rept. \textbf{561}, 1 (2014)
  [arXiv:1410.0679].

\bibitem{Andersen:2020hvz}
  B.~C.~Andersen {\it et al.},
  Nature {\bf 587}, no. 7832, 54 (2020)
  [arXiv:2005.10324].

\bibitem{Bochenek:2020zxn}
  C.~D.~Bochenek {\it et al.},
  Nature {\bf 587}, no. 7832, 59 (2020)
  [arXiv:2005.10828].

\bibitem{Lin:2020mpw}
  L.~Lin {\it et al.},
  Nature {\bf 587}, no. 7832, 63 (2020)
  [arXiv:2005.11479].

\bibitem{Li:2020qak}
  C.~K.~Li {\it et al.},
  Nat.\ Astron.\  {\bf 5}, 378 (2021)
  [arXiv:2005.11071].

\bibitem{Tendulkar:2017vuq}
  S.~P.~Tendulkar \textit{et al.},
  Astrophys. J. Lett. \textbf{834}, no.2, L7 (2017)
  [arXiv:1701.01100].

\bibitem{Marcote:2020ljw}
  B.~Marcote \textit{et al.},
  Nature \textbf{577}, no.7789, 190 (2020)
  [arXiv:2001.02222].

\bibitem{Niu:2021bnl}
  C.~H.~Niu \textit{et al.},
  Nature \textbf{606}, no.7916, 873 (2022)
  [arXiv:2110.07418].

\bibitem{Bhardwaj:2021hgc}
  M.~Bhardwaj \textit{et al.},
  Astrophys. J. Lett. \textbf{919}, no.2, L24 (2021)
  [arXiv:2108.12122].

\bibitem{Piro:2021upe}
  L.~Piro \textit{et al.},
  Astron. Astrophys. \textbf{656}, L15 (2021)
  [arXiv:2107.14339].

\bibitem{Xu:2021qdn}
  H.~Xu \textit{et al.},
  arXiv:2111.11764 [astro-ph.HE].

\bibitem{Bhardwaj:2021xaa}
  M.~Bhardwaj \textit{et al.},
  Astrophys. J. Lett. \textbf{910}, no.2, L18 (2021)
  [arXiv:2103.01295].

\bibitem{Kirsten:2021llv}
  F.~Kirsten \textit{et al.},
  Nature \textbf{602}, no.7898, 585 (2022)
  [arXiv:2105.11445].

\bibitem{Nimmo:2021yob}
  K.~Nimmo \textit{et al.},
  Nat.\ Astron.\  {\bf 6}, 393 (2022)
  [arXiv:2105.11446].

\bibitem{Zhang:2021kdu}
  R.~C.~Zhang and B.~Zhang,
  Astrophys. J. Lett. \textbf{924}, no.1, L14 (2022)
  [arXiv:2109.07558].

\bibitem{CHIMEFRB:2021srp}
  M.~Amiri \textit{et al.},
  Astrophys. J. Supp. \textbf{257}, no.2, 59 (2021)
  [arXiv:2106.04352].\\
  The data for CHIME/FRB Catalog 1 in machine-readable format can be
  found via their public webpage at https:$/\!/$www.chime-frb.ca/catalog

\bibitem{Qiang:2021ljr}
  D.~C.~Qiang, S.~L.~Li and H.~Wei,
  JCAP \textbf{2201}, 040 (2022)
  [arXiv:2111.07476].

\bibitem{Deng:2013aga}
  W.~Deng and B.~Zhang,
  Astrophys.\ J.\  {\bf 783}, L35 (2014)
  [arXiv:1401.0059].

\bibitem{Yang:2016zbm}
  Y.~P.~Yang and B.~Zhang,
  Astrophys.\ J.\  {\bf 830}, no. 2, L31 (2016)
  [arXiv:1608.08154].

\bibitem{Gao:2014iva}
  H.~Gao, Z.~Li and B.~Zhang,
  Astrophys.\ J.\  {\bf 788}, 189 (2014)
  [arXiv:1402.2498].

\bibitem{Zhou:2014yta}
  B.~Zhou, X.~Li, T.~Wang, Y.~Z.~Fan and D.~M.~Wei,
  Phys.\ Rev.\ D {\bf 89}, 107303 (2014)
  [arXiv:1401.2927].

\bibitem{Qiang:2019zrs}
  D.~C.~Qiang, H.~K.~Deng and H.~Wei,
  Class.\ Quant.\ Grav.\  {\bf 37}, 185022 (2020)
  [arXiv:1902.03580].

\bibitem{Qiang:2020vta}
  D.~C.~Qiang and H.~Wei,
  JCAP {\bf 2004}, 023 (2020)
  [arXiv:2002.10189].

\bibitem{Qiang:2021bwb}
  D.~C.~Qiang and H.~Wei,
  Phys. Rev. D \textbf{103}, 083536 (2021)
  [arXiv:2102.00579].

\bibitem{Dolag:2014bca}
  K.~Dolag \textit{et al.},
  Mon. Not. Roy. Astron. Soc. \textbf{451}, no.4, 4277 (2015)
  [arXiv:1412.4829].

\bibitem{Prochaska:2019mn}
  J.~X.~Prochaska and Y.~Zheng,
  Mon. Not. Roy. Astron. Soc. \textbf{485}, no.1, 648 (2019)
  [arXiv:1901.11051].

\bibitem{Shannon:2018}
  R.~M.~Shannon {\it et al.},
  Nature {\bf 562}, no. 7727, 386 (2018).

\bibitem{Prochaska:2019sci}
  J.~X.~Prochaska {\it et al.},
  Science\ {\bf 366}, no. 6462, 231 (2019)
  [arXiv:1909.11681].

\bibitem{Hashimoto:2019aqu}
  T.~Hashimoto {\it et al.},
  Mon.\ Not.\ Roy.\ Astron.\ Soc.\  {\bf 488}, no. 2, 1908 (2019)
  [arXiv:1907.11730].

\bibitem{Cordes:2002wz}
  J.~M.~Cordes and T.~J.~W.~Lazio,
  astro-ph/0207156.

\bibitem{Cordes:2003ik}
  J.~M.~Cordes and T.~J.~W.~Lazio,
  astro-ph/0301598.

\bibitem{pygedm}
  https:$/\!/$pypi.org/project/pygedm

\bibitem{Aghanim:2018eyx}
  N.~Aghanim {\it et al.},
  Astron.\ Astrophys.\  {\bf 641}, A6 (2020)
  [arXiv:1807.06209].

\bibitem{Zhang:2018csb}
  B.~Zhang,
  Astrophys. J. Lett. \textbf{867}, no.2, L21 (2018)
  [arXiv:1808.05277].

\bibitem{Zhang:2020ass}
  R.~C.~Zhang {\it et al.},
  Mon. Not. Roy. Astron. Soc. \textbf{501}, 157 (2021)
  [arXiv:2011.06151].

\bibitem{Pietka:2014wra}
  M.~Pietka, R.~P.~Fender and E.~F.~Keane,
  Mon. Not. Roy. Astron. Soc. \textbf{446}, 3687 (2015)
  [arXiv:1411.1067].

\bibitem{Majid:2021uli}
  W.~A.~Majid {\it et al.},
  Astrophys. J. Lett. \textbf{919}, no.1, L6 (2021)
  [arXiv:2105.10987].

\bibitem{KStest}
  https:$/\!/$en.wikipedia.org/wiki/Kolmogorov-Smirnov$_{-}$test

\bibitem{KStestpy}
  https:$/\!/$docs.scipy.org/doc/scipy/reference/generated/scipy.stats.kstest.html

\bibitem{Chawla:2021igg}
  P.~Chawla \textit{et al.},
  Astrophys. J. \textbf{927}, 35 (2022)
  [arXiv:2107.10858].

\bibitem{Madau:2016jbv}
  P.~Madau and T.~Fragos,
  Astrophys. J. \textbf{840}, no.1, 39 (2017)
  [arXiv:1606.07887].

\bibitem{LinearRegression}
 https:$/\!/$scikit-learn.org/stable/modules/generated/sklearn.linear$_{-}$model.LinearRegression.html

\bibitem{Nimmo:2021ntn}
  K.~Nimmo \textit{et al.},
  Astrophys. J. Lett. \textbf{927}, no.1, L3 (2022)
  [arXiv:2111.01600].

\bibitem{Lu:2021enm}
  W.~Lu, P.~Beniamini and P.~Kumar,
  Mon. Not. Roy. Astron. Soc. \textbf{510}, 1867 (2022)
  [arXiv:2107.04059].

\bibitem{Luo:2020wfx}
  R.~Luo \textit{et al.},
  Mon. Not. Roy. Astron. Soc. \textbf{494}, 665 (2020)
  [arXiv:2003.04848].

\bibitem{Schechter:1976iz}
  P.~Schechter,
  Astrophys. J. \textbf{203}, 297 (1976).

\bibitem{Lu:2019pdn}
  W.~Lu and A.~L.~Piro,
  Astrophys. J. \textbf{883}, 40 (2019)
  [arXiv:1903.00014].

\bibitem{Luo:2018tiy}
  R.~Luo {\it et al.},
  Mon. Not. Roy. Astron. Soc. \textbf{481}, 2320 (2018)
  [arXiv:1808.09929].

\bibitem{McQuinn:2013tmc}
  M.~McQuinn,
  Astrophys. J. Lett. \textbf{780}, L33 (2014)
  [arXiv:1309.4451].

\bibitem{Ioka:2003fr}
  K.~Ioka,
  Astrophys. J. Lett. \textbf{598}, L79 (2003)
  [astro-ph/0309200].

\bibitem{Jaroszynski:2018vgh}
  M.~Jaroszynski,
  Mon. Not. Roy. Astron. Soc. \textbf{484}, no.2, 1637 (2019)
  [arXiv:1812.11936].

\end{thebibliography}
\end{document}